\documentclass[twocolumn, usenatbib]{mnras}
\pdfoutput=1 
\usepackage{amsmath,amstext}
\usepackage[T1]{fontenc}
\usepackage{ae,aecompl}
\usepackage[figure,figure*]{hypcap}
\usepackage{graphicx}
\usepackage{amssymb}	
\usepackage[caption=false]{subfig}
\newcommand{\angstrom}{\textup{\AA}}

\makeatletter
\renewcommand*{\@fnsymbol}[1]{\ifcase#1\or ^\dagger\or ^{\star} \else\@ctrerr\fi}
\makeatother
\title[Strongly Lensed Galaxies Discovered in MaNGA]{SDSS-IV MaNGA: The Spectroscopic Discovery of Strongly Lensed Galaxies}

\author[Talbot and Brownstein et al.]{
\parbox{\textwidth}{Michael S.\,Talbot,$^{1}$\thanks{E-mail: \texttt{\href{mailto:michaeltalbot@astro.utah.edu}{michaeltalbot@astro.utah.edu}}}
Joel R.\,Brownstein,$^{1}$\thanks{E-mail: \texttt{\href{mailto:joelbrownstein@physics.utah.edu}{joelbrownstein@physics.utah.edu}}}
Adam S.\,Bolton,$^{2}$
Kevin Bundy,$^{3}$
Brett H.\,Andrews,$^{4}$
Brian Cherinka,$^{5}$
Thomas E.\,Collett,$^{6}$
Anupreeta More,$^{7}$
Surhud More,$^{7}$
Alessandro Sonnenfeld,$^{7}$
Simona Vegetti,$^{8}$
David A.\,Wake,$^{9,10}$
Anne-Marie Weijmans,$^{11}$
Kyle B.\,Westfall$^{3}$\\}
\\
$^{1}$Department of Physics and Astronomy, University of Utah, 115 S. 1400 E., Salt Lake City, UT 84112, USA \\
$^{2}$National Optical Astronomy Observatory 950 North Cherry Avenue, Tucson, AZ 85719, USA \\
$^{3}$UCO/Lick Observatory, University of California, Santa Cruz, 1156 High St. Santa Cruz, CA 95064, USA \\
$^{4}$PITT PACC, Department of Physics and Astronomy, University of Pittsburgh, Pittsburgh, PA 15260, USA \\
$^{5}$Center for Astrophysical Sciences, Department of Physics and Astronomy, Johns Hopkins University, 3400 North Charles Street,\\ \quad Baltimore, MD 21218, USA \\
$^{6}$Institute of Cosmology \& Gravitation, University of Portsmouth, Dennis Sciama Building, Portsmouth, PO1 3FX, UK \\
$^{7}$Kavli IPMU (WPI), UTIAS, The University of Tokyo, Kashiwa, Chiba 277-8583, Japan \\
$^{8}$Max-Planck-Institut f\"ur Astrophysik, Karl-Schwarzschild-Str. 1, D-85748 Garching, Germany \\
$^{9}$School of Physical Sciences, The Open University, Milton Keynes MK7 6AA, UK \\
$^{10}$Department of Physics, University of North Carolina, Asheville, NC 28804, USA \\
$^{11}$School of Physics and Astronomy, University of St Andrews, North Haugh, St Andrews KY16 9SS, UK
}

\date{Accepted 2018 March 08. Received 2018 March 08; in original form 2018 January 04}
\pubyear{2018}
\begin{document}
\maketitle
\begin{abstract}

We present a catalogue of 38 spectroscopically detected strong galaxy-galaxy gravitational lens candidates identified in the Sloan Digital Sky Survey IV (SDSS-IV).  We were able to simulate narrow-band images for 8 of them demonstrating evidence of multiple images. Two of our systems are compound lens candidates, each with 2 background source-planes.  One of these compound systems shows clear lensing features in the narrow-band image.  Our sample is based on 2812 galaxies observed by the Mapping Nearby Galaxies at APO (MaNGA) integral field unit (IFU).  This Spectroscopic Identification of Lensing Objects (SILO) survey extends the methodology of the Sloan Lens ACS Survey (SLACS) and BOSS Emission-Line Survey (BELLS) to lower redshift and  multiple IFU spectra.  We searched $\sim 1.5$ million spectra, of which 3065 contained multiple high signal-to-noise background emission-lines or a resolved [O\,{\sc ii}]\ doublet, that are included in this catalogue. Upon manual inspection, we discovered regions with multiple spectra containing background emission-lines at the same redshift, providing evidence of a common source-plane geometry which was not possible in previous SLACS and BELLS discovery programs.  We estimate more than half of our candidates have an Einstein radius \(\gtrsim 1.7\arcsec\), which is significantly greater than seen in SLACS and BELLS.  These larger Einstein radii produce more extended images of the background galaxy increasing the probability that a background emission-line will enter one of the IFU spectroscopic fibres, making detection more likely.

\end{abstract}

\section{Introduction}
\label{section:introduction}

Galaxy-scale strong gravitational lensing provides a unique probe into the cosmological distribution of matter, providing precision measurements of foreground galaxy surface mass densities from high-resolution lens models of observed multiple images~\citep{1995ApJ...445..559K, 1998ApJ...509..561K,2008ApJ...682..964B,2008ApJ...684..248B,2009ApJ...705.1099A,2010ApJ...724..511A,2012ApJ...744...41B,2012MNRAS.424.2800L,2013ApJ...777...97S,2016ApJ...833..264S}.   When combined with other dynamical measurements, such as stellar kinematics, the mass distribution may be dissected into luminous and dark components~\citep{2009MNRAS.399...21B,2011MNRAS.417.1621D,2011MNRAS.417.3000S,2012MNRAS.423.1073B}, including the detection of dark matter substructure within the lensing galaxies as well as small-mass dark matter haloes along their line-of-sight \citep{Vegetti09a, 2010MNRAS.408.1969V,2012Natur.481..341V,2014MNRAS.442.2017V,2016ApJ...823...37H,2018ApJ...853..148C,2018MNRAS.475.5424D}.

Dedicated surveys and serendipitous discoveries have to-date identified over 250 grade-A strong galaxy-galaxy gravitational lenses\footnote{L.~Moustakas and J.~R. Brownstein: Master Lens Database}, with foreground galaxy subtractions which exhibit multiple images of the same background galaxy with sufficient signal-to-noise to lead to a well-constrained lens model. Of the various methods, spectroscopic detection has proven to be the most successful means of discovery, with the Sloan Lens ACS~\citep[SLACS;][]{2006ApJ...638..703B} survey, the Sloan WFC Edge-on Late-type Lens Survey~\citep[SWELLS;][]{2011MNRAS.417.1601T}, the SLACS for the Masses~\citep[S4TM;][]{2015ApJ...803...71S} survey, the BOSS Emission-Line Lens Survey~\citep[BELLS;][]{2012ApJ...744...41B} and the BELLS Galaxy-Ly$\alpha$ Emitter systems~\citep[GALLERY;][]{2016ApJ...824...86S} survey yielding a combined total of over 150 grade-A strong galaxy-galaxy lenses, which were detected from the Sloan Digital Sky Survey~\citep{2000AJ....120.1579Y}.  All SDSS spectroscopic candidates required follow-up high-resolution Hubble Space Telescope (HST) imaging for confirmation of multiple background source images.

This manuscript introduces a new Spectroscopic Identification of Lensing Objects (SILO) survey based on the current SDSS-IV survey~\citep{2017AJ....154...28B}, extending the spectroscopic detection methods of \cite{2004AJ....127.1860B} and \cite{2012ApJ...744...41B} to spectra obtained from the Mapping of Nearby Galaxies at APO~\citep[MaNGA;][]{2015ApJ...798....7B} survey.   Unlike BOSS and SDSS-I spectra, which were reduced from single fibres placed on each galaxy, MaNGA's Integral Field Unit~\citep[IFU;][]{2015AJ....149...77D, 2016AJ....151....8Y} places between 19 and 127  $2\arcsec$ fibres on each galaxy, covering a much larger field of view, between $12\arcsec$ -- $32\arcsec$ in diameter, with 15 minute dithered exposures repeated until the signal-to-noise ratio above a threshold is achieved \citep{2016AJ....152..197Y}.  MaNGA's six-year program targets $\sim 10,000$ nearby ($z < 0.15$) galaxies~\citep{2015AJ....150...19L} of which 2,812 galaxies were included in Data Release 14~\citep[DR14;][]{2017arXiv170709322A}.

Whereas SLACS, SWELLS and S4TM surveys were based on SDSS-I galaxies with redshifts out to 0.45~\citep{2000AJ....120.1579Y, 2001AJ....122.2267E} and BELLS and BELLS GALLERY surveys were based on SDSS-III BOSS galaxies with redshifts out to 0.7~\citep{2011AJ....142...72E, 2013AJ....145...10D}, the typical MaNGA target redshift is \(z \sim 0.05\).  The majority of the MaNGA targets are selected using $i$-band absolute magnitude (M$_i$)-dependent redshift cuts such that more luminous galaxies are selected at both preferentially higher redshift and over a larger volume than lower luminosity galaxies. The first of theses choices results in targets that cover the same range in angular effective radius ($R_\mathrm{eff}$) irrespective of M$_i$, and so all galaxies can be sampled to some multiple of $R_\mathrm{eff}$ using the same size IFUs. The second produces a sample that has the same number density at all M$_i$, i.e. the sample has the same number of the most luminous (massive) galaxies as the least luminous (massive) galaxies. Two main samples are defined. The Primary+ sample is selected so that galaxies have spectral coverage out to 1.5 $R_\mathrm{eff}$ and makes up two thirds of the targets. The Secondary sample, making up the remaining third of the targets, is selected at slightly higher redshift than the Primary sample so that the galaxies have spectral coverage out to 2.5 $R_\mathrm{eff}$. Full details of the MaNGA target selection can be found in \citet{2017AJ....154...86W}.

The discoveries presented in this manuscript are based on the MaNGA internal {\sc MPL-5} data release which contains the same galaxies as in DR14. Whereas we have fewer galaxies in MaNGA than in the earlier lensing spectroscopic discovery programs, we have many more spectra for each individual galaxy, allowing the SILO detection algorithm to search for multiple background emission-lines with common background source redshifts.   Furthermore, because the mean redshift of MaNGA (foreground) galaxies is $z \sim 0.05$ with background galaxy redshifts up to the range of the previous surveys, the estimated Einstein radius is up to two to three times larger than those measured in the SLACS, SWELLS, S4TM, BELLS and BELLS GALLERY surveys.  Such large Einstein radii would not have been possible to detect with single fibre spectra because the background emission-lines of the magnified image of the source would have gone undetected since it would have landed outside the single fibre.

The larger Einstein radius is a measurement of the increased gravitational lensing strength, which produces more extended images of the background galaxy, thereby increasing the probability that a background emission-line will enter one of the IFU spectroscopic fibres, making detection more likely.  This increases the overall probability that a galaxy of a given mass will be a strong gravitational lens. The multiple fibres with multiple exposures within the Einstein radius allow a correlation of the spectroscopic detections with their geometrical position across the IFU bundle.  Using the individual best-fit spectroscopic redshifts of detected background sources, we have extended the SILO algorithm to generate narrow-band images of the background source-plane, to allow for the first time, experimental predictions of images prior to follow-up imaging.

Shortly after the first MaNGA data release in 2016~\citep[DR13;][]{2017ApJS..233...25A}, which included 1,351 galaxies, \cite{2017MNRAS.464L..46S} reported the discovery of a strong-lensing candidate, SDSS~J170124.01+372258, at $z_\mathrm{F} = 0.122$.  We also find this system as a promising lens candidate, based on our discovery of high signal-to-noise background emission-lines from a source at $z_\mathrm{B} = 0.790$. The multiple images seen in the narrow-band images are geometrically positioned at the estimated Einstein radius and clearly show an arc and a counter-image separated by $~ 3.7\pm1.3\arcsec$.

This paper is organized as follows.  The expected number of lenses in MaNGA is presented in Section~\ref{section:expectation}.  The spectroscopic candidate discovery is presented in Section~\ref{section:discovery}, consisting of foreground spectrum modelling (Section~\ref{subsection:discovery.galaxy_template}), background emission-line detection (Section~\ref{subsection:discovery.emline}), Einstein radius estimation (Section~\ref{subsection:discovery.einstein}), and source-plane inspection, with the discovery catalogue resulting from inspection (Section~\ref{subsection:discovery.inspection}).  Lensed image construction,  consisting of the discovery of counter-images in narrow-band images, is presented in  Section~\ref{section:lensing}.  Potential compound lenses containing two background galaxies along the same line of sight which were discovered in the catalogue are discussed in Section~\ref{section:compound}.  We provide a summary of results and conclusions on the importance of the SILO survey's MaNGA lens discovery program in Section~\ref{section:conclusions}.

\section{Expected number of lenses}
\label{section:expectation}

We begin by presenting a crude estimate for the number of gravitational lenses expected to be detected in MaNGA. We make several simplifying assumptions. For instance, we assume that the lenses will be detected if the [O{\sc ii}] luminosity in the source is above a certain flux limit and we ignore the magnification boost provided by strong lensing. The probability that a given MaNGA galaxy with velocity dispersion $\sigma_v$ at a redshift $z_{\mathrm L}$ will act as a lens for a source galaxy with flux brighter than $f$ is then given by
\begin{equation}
P(z_{\mathrm L}, \sigma_v, f) = \int n_{\mathrm S}(z_{\mathrm S}, z_{\mathrm L}, f)\,\pi \theta_\mathrm {Ein}^2(\sigma_v, z_{\mathrm L}, z_{\mathrm S}) \frac{dV}{dz_{\mathrm S}} dz_{\mathrm S}
\end{equation}
where $n_{\mathrm S}$ is the co-moving source number density of [O{\sc ii}] galaxies at redshift $z_{\mathrm S}$ with an observed flux greater than $f$ (which translates to a luminosity $L>4\pi D_{\rm L}^2 f$ given the luminosity distance $D_{\rm L}$), $\theta_\mathrm{Ein}$ is the Einstein radius, given the lens and source redshifts, and $dV/d z_{\mathrm S}$ is the differential co-moving volume at redshift $z_{\mathrm S}$. We use the Schechter function fits to the [O{\sc ii}] luminosity function and its evolution \citep{2015MNRAS.452.3948K} to compute $n_{\mathrm S}$ and interpolate the redshift ranges in between.  The Einstein radius is then given by
\begin{equation}
\theta_\mathrm{Ein} = 4\pi \frac{\sigma_{v}^2}{c^2} \frac{D_{\rm A}(z_{\mathrm L}, z_{\mathrm S})}{D_{\rm A}(z_{\mathrm S})}
\end{equation}
where, $c$ is the speed of light, and $D_{\rm A}(z_{\mathrm L}, z_{\mathrm S})$ and $D_{\rm A}(z_{\mathrm S})$ are the angular diameter distances between the lens and the source, and the source and observer, respectively.  We used a fiducial value of \(\sigma_{v} \sim 200\) km/s for this coarse estimate.

If we assume the [O{\sc ii}] flux limit to be $6\times 10^{-17} {\rm erg\,s^{-1} cm^{-2}}{\angstrom}^{-1}$, we arrive at an estimate of $\sim 60$ lenses in the entire MaNGA survey. This estimate increases to $\sim 200$ lenses if the flux limit is reduced to $10^{-17} {\rm erg\,s^{-1} cm^{-2}}{\angstrom}^{-1}$. We have checked that accounting for the magnification bias changes the prediction by a factor of order unity.  A fuller investigation into the biases of the lensing probability~\citep[e.g.,][]{2012ApJ...753....4A} is in preparation.
\section{Spectroscopic Candidate Discovery}\label{section:discovery}

The premise behind the spectroscopic  candidate selection of \cite{2004AJ....127.1860B} and \cite{2012ApJ...744...41B} was to search for background emission-lines within the solid angle covered by the spectroscopic fibre \cite[see also][]{1996MNRAS.278..139W, 2000ASPC..195...94H, 2005MNRAS.363.1369W, 2006MNRAS.369.1521W}.

We first summarize our approach for the lens search here and then describe each step in detail subsequently. The success of this discovery program hinges on a precise subtraction of a best-fit foreground galaxy model.  
Although there are many residual features remaining in the spectrum after foreground galaxy subtraction, we limit our search for background emission-lines to high signal-to-noise features. 
We then use an upper limit on the Einstein radius for each galaxy, in order to determine which of our background emission-line regions were more likely strongly lensed images. Furthermore, because the background source-plane would occur at a common background redshift, and images of the background galaxy would be localized in a particular geometrical position in the IFU, a manual inspection process is required in order to grade the candidates for the final catalogue. 

\subsection{Foreground Galaxy Subtraction}\label{subsection:discovery.galaxy_template}

\begin{figure}
    \centering
    \includegraphics[width=.5\textwidth]{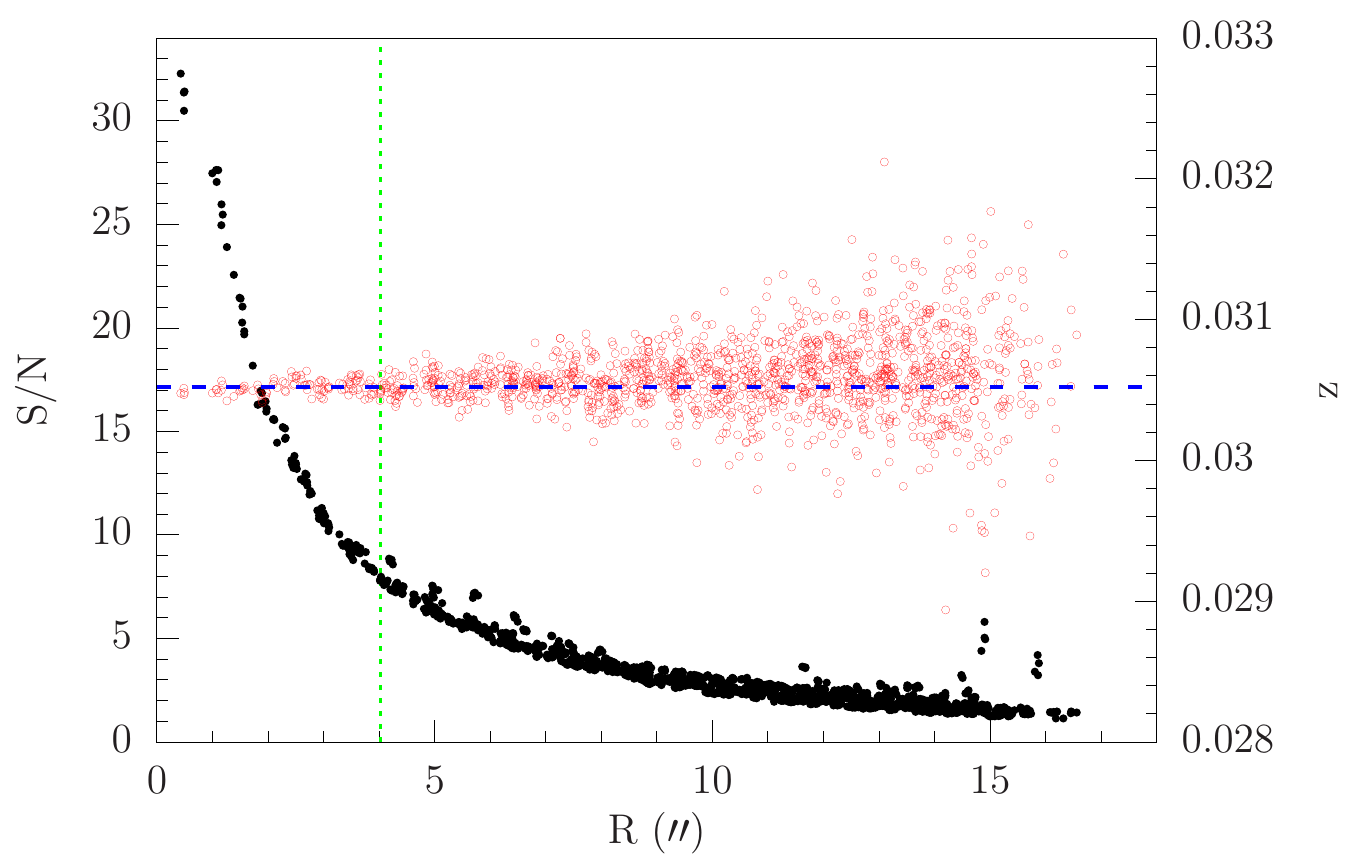}
    \caption{The distribution of the mean signal-to-noise (S/N), shown in black solid-circles with respect to left vertical axis, and best-fit spectroscopic redshift, shown in open red-circles with respect to the right vertical axis (as explained in Section~\ref{subsection:discovery.galaxy_template}) for each spectrum, plotted as a function of the mean radial position (in arc-seconds) from the centre of the galaxy, shown for SDSS~J0836+5402. The high S/N region is inside the mean inner radius, plotted as a vertical green-dotted line. The mean spectroscopic redshift is shown by the horizontal blued-dashed line.}
    \label{figure:specz}
\end{figure}

The fact that MaNGA galaxies are observed using the BOSS Spectrograph~\citep{2006AJ....131.2332G, 2013AJ....146...32S} and that the row-stacked spectra (RSS) files generated by the MaNGA Data Reduction Pipeline~\cite[DRP;][]{2016AJ....152...83L} are equivalent to the BOSS and SDSS-I/II {\sc spec2d} reductions allows a very similar foreground galaxy model as was used in the SLACS~\citep{2006ApJ...638..703B}, SWELLS~\citep{2011MNRAS.417.1601T}, S4TM~\citep{2015ApJ...803...71S}, BELLS~\citep{2012ApJ...744...41B} and GALLERY~\citep{2016ApJ...824...86S} spectroscopic discovery programs. Whereas the BOSS and SDSS-I galaxies were observed with one fibre yielding a single spectrum containing all of the light from the galaxy, the MaNGA RSS files contain multiple spectra for each fibre, repeated across multiple exposures, yielding many spectra distributed at different positions over the galaxy.  The wavelength calibration of the MaNGA data is accurate to 5 km\,s$^{-1}$ RMS, with a median spatial resolution of \(2.54^{\prime \prime}\) FWHM. This allows the candidate background emission-lines to be spatially correlated, increasing our confidence that the background emission-lines are real.
\begin{table}
\caption{\label{table:emline}Emission-Line Wavelengths. Column 1 lists the emission-lines used in the spectroscopic discovery of background emission-line sources, and Column 2 shows the restframe vacuum wavelength of the emission-line.  Column 3 provides the maximum redshift of the emission-line detectable with the BOSS spectrograph used by MaNGA.}
\centering
\begin{tabular}{ c c c }
\hline
\hline
{Emission} & {Restframe} & {\(z_{\mathrm{max}}\)} \\
{Line} & {Wavelength [\AA]} &  \\
{\scriptsize (1)} & {\scriptsize (2)} & {\scriptsize (3)} \\
\hline
O\,{\sc ii} & 3727.09 & 1.78 \\
O\,{\sc ii} & 3729.88 & 1.78 \\
H${\delta}$ & 4102.89 & 1.52 \\
H${\gamma}$ & 4341.68 & 1.38 \\
H${\beta}$ & 4862.68 & 1.13 \\
O\,{\sc iii} & 4960.30 & 1.09 \\
O\,{\sc iii} & 5008.24 & 1.07 \\
N\,{\sc ii} & 6549.86 & 0.58 \\
H${\alpha}$ & 6564.61 & 0.58 \\
N\,{\sc ii} & 6585.27 & 0.57 \\
S\,{\sc ii} & 6718.29 & 0.54 \\
S\,{\sc ii} & 6732.68 & 0.54 \\

\hline
\end{tabular}
\end{table}
\newcommand{\fluxcaption}{
Example plots of typical multi-line and single-line detections. The black solid-line shows the MaNGA RSS observed flux density, \(f_\lambda\), as a function of observed (lower axis) and rest-frame (upper axis) wavelengths. The blue dashed-line shows the model fitted to the continuum of the foreground galaxy, and the red vertical dashed-line shows the wavelength of the discovered background emission-lines.  The green dashed-dotted line shows the Gaussian fits to the background emission-lines.
}
\begin{figure*}\setcounter{subfigure}{1}
\begin{center}
\label{figure.flux}
\includegraphics[page=1, width=1\textwidth]{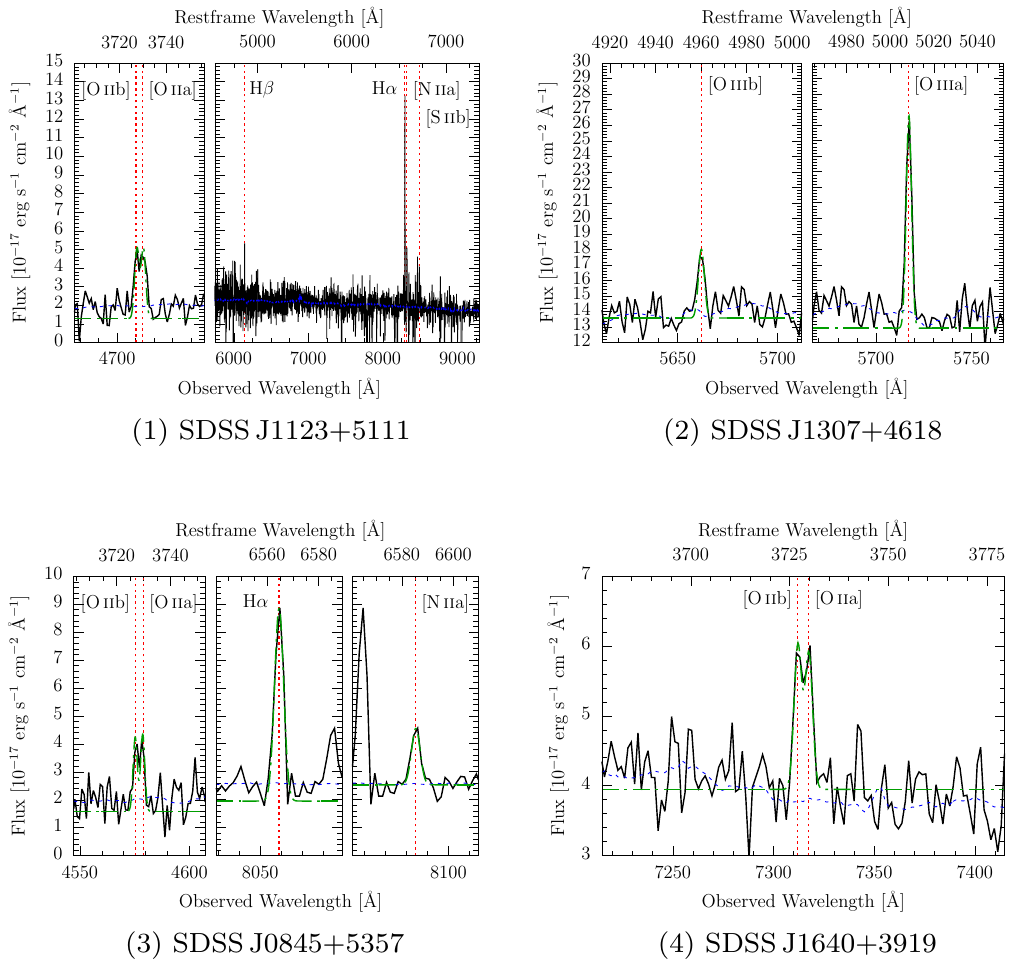}
\end{center}
\caption{\fluxcaption}
\end{figure*}

The galaxy model-fitting template requires a precise foreground redshift per fibre. Unlike the BOSS and SDSS-I/II reductions, the MaNGA DRP does not provide spectroscopic redshifts for each fibre.  The MaNGA targeting NASA-Sloan Atlas \cite[NSA;][]{2017ApJS..233...25A} catalogue redshift is insufficiently precise to match each fibre's redshift due to the variations caused by the galaxy's radial velocity distribution.  For each row (spectrum) in the RSS file, we computed the  best-fit spectroscopic redshift by running the publicly available BOSS pipeline {\sc spec1d -- zfind} code~\citep{2012AJ....144..144B}.  Because of the variation in the signal-to-noise which decreases rapidly from the centre of the galaxy, we needed to make two passes to get good spectroscopic redshifts everywhere possible.  The first pass computed spectroscopic redshifts using the provided value of the NSA catalogue redshift as the starting guess (prior) input to {\sc spec1d -- zfind}.  A mean spectroscopic redshift, \({\bar z_{\mathrm{spec}}}\), was then computed within the inner high signal-to-noise region, determined as the minimum radius at which the curvature became slowly varying as defined by \(|d^2(S/N)/dR^2| \rightarrow 1 \), where \(S/N\) is the signal-to-noise ratio.  Redshifts that were more than 500 km/s from the mean were neglected as outliers.  The second pass re-computed the spectroscopic redshifts in the low signal-to-noise region using the mean spectroscopic redshift as a prior.  This method provided good spectroscopic redshifts throughout the galaxy, except in a minority of fibres, as shown in Figure~\ref{figure:specz}, for an example galaxy, SDSS~J083633.11+540229.27, with the mean spectroscopic z, shown by the blue-dashed line, very close to the NSA catalogue redshift, 
\begin{equation}
{\bar z} = 0.03052 \sim z_{nsa} = 0.03046.
\end{equation}

In total, we obtained 1,582,014 good spectroscopic redshifts for the spectra of the 2,812 galaxies in MaNGA's internal {\sc MPL-5} data release.  The mean inner spectroscopic redshift agreed with the NSA catalogue redshift to within \(0.001\) for 2,750 of the 2,812 galaxies.  For each of these spectra, we constructed a best-fit model spectrum to the galaxy continuum using a basis of 7 principal component analysis (PCA) eigenspectra, and subtracted this continuum model from the data.
Fibres which were at too low a S/N to yield a spectroscopic redshift were discarded from consideration in the search for background emission-lines.

\subsection{Background Emission-Line Detection}\label{subsection:discovery.emline}

In order to perform a systematic search for background emission-lines in each of the \(\gtrsim 1.5M\) residual spectra described in Section~\ref{subsection:discovery.galaxy_template}, we reimplemented the original {\sc idl} based algorithms described in \cite{2004AJ....127.1860B} and \cite{2012ApJ...744...41B} in a python-based software architecture including a back-end database, and utilized the {\sc marvin} suite \citep{zenodo.292632} to access MaNGA data including the web to visualize spectra and galaxy maps.   Using a 500 cpu cluster, the residual spectra were scanned for at least two of the lines listed in Table~\ref{table:emline} detected at \(4\sigma\); or a single line  identified as a background [O\,\textsc{ii}]\(\lambda\lambda\)\,3727 detected at \(6\sigma\).  These are the values of the thresholds used in \cite{2004AJ....127.1860B} and \cite{2012ApJ...744...41B} for these two detection modes, referred to as ``multi-line'' and ``single-line'' detections, respectively.   Decreasing the threshold increases the number of candidate emission-line detections without necessarily increasing the number of candidates that pass inspection, whereas increasing the values of these thresholds decreases the number of candidates.

Detections that contained only an [O\,\textsc{ii}] doublet were rejected if the [O\,\textsc{ii}] doublet was located in a region near a sky emission-line. Multi-line detections were modelled using a Gaussian, and detections that included both [O\,\textsc{iii}]\,5007 and [O\,\textsc{iii}]\,4959 emission-lines were scanned for a characteristic $\sim$ 3-to-1 ratio of the flux, respectively. All three multi-line or greater hits were also visually inspected. Single-line detections were modelled with double Gaussian functions in order to scan for resolved [O\,\textsc{ii}] doublets. We inspected 1659 multi-line ``hits'' and 9762 single-line ``hits'' that contained either the form of an [O\,\textsc{ii}] doublet, or contained multiple patterns of background emission-lines. Each of these hits were graded as good. However, a large fraction of the single-line hits within a particular galaxy are not correlated in either redshift or spatially across the IFU and therefore drop out during the source-plane inspection process. Furthermore, detections that were geometrically positioned beyond the estimated strong lensing regime (see Section \ref{subsection:discovery.inspection}) were discarded as non-candidates.

We provide example plots of some of the multi-line and single-line detections in Figure~\ref{figure.flux}, showing the flux, best-fit model, and Gaussian fits, including the observed emission-lines, as described in Table~\ref{table:emline_inspection}.

\begin{table*}
\caption{\label{table:emline_inspection}Background Emission Line Inspection. This list identifies the background emission-lines that are observed within the examples presented in Figure~\ref{figure.flux}. Column 1 presents the SDSS System Name. Column 2 identifies the detection mode used for the discovery. Column 3 provides the manual inspection which identifies the observed emission-lines.}
\centering
\begin{tabular}{ l l l }
\hline
\hline
{System Name} & {Detection Mode} & {Manual Inspection} \\
\phantom{Syste }{\scriptsize (1)} & \phantom{Detecti}{\scriptsize (2)} & \phantom{Manual }{\scriptsize (3)}\\
\hline
SDSS~J1123+5111 & {Multi-line} & \parbox{12cm}{Resolved [O\,\textsc{ii}] doublet, H$\alpha$, H$\beta$, [N\,\textsc{ii}a], and [S\,\textsc{ii}b].} \\
SDSS~J1307+4618 & {Multi-line} & {Strong [O\,\textsc{iii}a] and [O\,\textsc{iii}b] with a characteristic 3:1 ratio.} \\
SDSS~J0845+5357 & {Multi-line} & {Resolved [O\,\textsc{ii}] doublet, H$\alpha$, and N\,\textsc{ii}a.} \\
SDSS~J1640+3919 & {Single-line} & {Resolved [O\,\textsc{ii}] doublet.} \\
\hline
\end{tabular}
\end{table*}

As in previous discovery programs (including SLACS and BELLS), these background emission-line detections provide compelling evidence that follow-up imaging will confirm the existence of strong gravitational lenses.  But we can go farther in our inspection than was previously possible with the single spectroscopic observation of SDSS and BOSS, by combining an estimate of the Einstein radius, as described in Section~\ref{subsection:discovery.einstein}, with the geometric information with a source-plane inspection, as described in Section~\ref{subsection:discovery.inspection}.  

\subsection{Einstein Radius Estimation}\label{subsection:discovery.einstein}
MaNGA data allows us to use spatial information to classify our candidate lens systems. For strong lenses with a density profile close to isothermal, such as SLACS lenses \citep{2009ApJ...703L..51K}, we expect strongly lensed images to lie within a circular region with radius equal to twice the Einstein radius of the lens. Although the most accurate measurement of the Einstein radius is derived by fitting a lens model to high-resolution imaging data, we do not have a complete set of follow-up imaging for our catalogue to model. Instead, we estimate the Einstein radius, using:
\begin{equation} \label{equation:einstein_r}
    R_{\mathrm{Ein}} = \sqrt {\frac{4GM}{c^2}\frac{D_{LS}}{D_L D_S}}
\end{equation}
 where \(M\) is the enclosed mass, and \(D_{L}\) and \(D_S\) are cosmological angular diameter distances from observer to the lens and observer to the source, respectively.

The computation of the Einstein radius requires surface mass density maps for the galaxy, which we obtained from the publicly available MaNGA Firefly value-added catalogue \citep[VAC,][]{2015MNRAS.449..328W} released with SDSS-IV DR14~\citep{2017arXiv170709322A}.  In this VAC the input stellar population models of \citet{2011MNRAS.418.2785M} are used, based on the MILES stellar library, with Kroupa IMF~\citep{2001MNRAS.322..231K}.  We computed both a lower limit and an upper limit for the Einstein radius. For the lower limit of the Einstein radius, we computed the stellar mass enclosed within the lower-bound of the error provided by Firefly, but added no dark matter.  For the upper limit, we computed the enclosed stellar mass within the upper-bound of the error provided by Firefly, and included a dark matter fraction computed using the theoretical model for the local stellar mass fraction of \cite{2015ApJ...799..149J} at values of the ratio \(R/R_{\mathrm{eff}}\), where \(R_{\mathrm{eff}}\) is the effective radius from the NSA catalogue.  We computed the mass, \(M\), enclosed within a distance, \(R\), and the Einstein radius, \(R_{\mathrm{Ein}}\), iteratively using Equation~\ref{equation:einstein_r}, until we obtained a \(R = R_{\mathrm{Ein}}\) solution.  

\begin{figure}
    \centering
    \includegraphics[width=.5\textwidth]{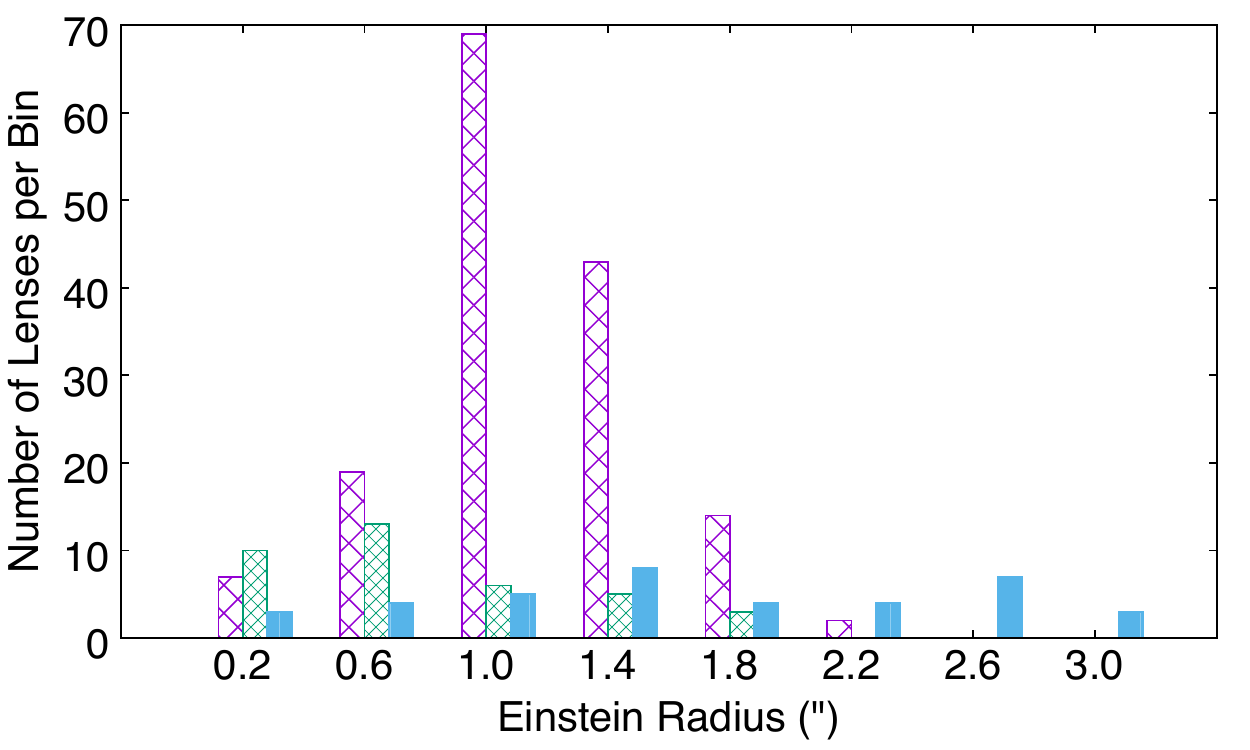}
    \caption{Distribution of the Einstein radii. The grade A, B, and C lenses from the SLACS (large hatched region in purple) and the BELLS (small hatched region in green) surveys are compared with the Einstein radii of the lens candidates from the SILO survey (solid blue).}
    \label{figure:erhistogram}
\end{figure}

We compared the distribution of the Einstein radii of the SILO candidates to the published distribution of the Einstein radii of the grade A, B, and C lenses in the SLACS and BELLS surveys. As displayed in Figure~\ref{figure:erhistogram}, the Einstein radius distributions of the SLACS and BELLS surveys peak at $\sim0.6\arcsec$ and $\sim1.0\arcsec$, with upper limits up to $\sim1.8\arcsec$ and $\sim2.2\arcsec$, respectively. The distribution of the Einstein radii for the SILO candidates peaks at $\sim1.4\arcsec$ and $\sim2.6\arcsec$, and the range extends up to $\sim3\arcsec$.

\subsection{Source-Plane Inspection}\label{subsection:discovery.inspection}

Whereas the BOSS and SDSS-I galaxies were observed with one fibre yielding a single spectrum containing all of the light from the galaxy, each MaNGA galaxy was observed with a fibre bundle, in which each fibre yielded multiple spectra from multiple exposures. This allows the candidate background emission-lines to be spatially correlated, increasing our confidence that the background emission-lines are real.

In order to determine whether the individual background emission-line hits discussed in Section~\ref{subsection:discovery.emline} originated in a common background source-plane, a finely spaced redshift grid (\(\Delta z = 0.05\)) was constructed and the hits for each galaxy were counted per spacing in the redshift grid. The background redshift, \(z_B\), was selected from one of the good hits, limited to a precision of 3 decimal places, to allow for the variation due to internal radial motion of the background galaxy. Each MaNGA galaxy that contained at least two good hits in a source-plane was manually inspected to check for spatial correlation in the emission-line features. Since the background emission-lines in one hit can often be observed in surrounding fibres, we also inspected the source-planes that contained one good hit with three or more multi-lines, or one good single-line hit within a MaNGA galaxy that contained another source-plane with at least two good hits. These search methods identified a total of 1139 good multi-line and 1926 good single-line hits, for a total of 3065. We inspected 397 background source-planes that contained at least two good hits. We inspected an additional 44 source-planes with a single multi-line hit and 257 source-planes with a single-line hit. 

\begin{table*}
\caption{\label{table:source_inspection}Candidate Strong Gravitational Lenses. This is the list of strong gravitational lens candidates, discovered in Section~\ref{subsection:discovery.emline} that further demonstrated background emission-lines at a common redshift (source-plane) within the strong lensing regime as discussed in Section~\ref{subsection:discovery.inspection}.  Column 1 provides the SDSS System Name in terms of truncated J2000 RA and Dec in the format HHMMSS.ss\(\pm\)DDMMSS.s.  A dagger represents if the candidate is displayed as a narrow-band image within this paper. Column 2 provides the Plate-IFU of the galaxy observation.  Column 3 provides the foreground spectroscopic redshift defined as the mean inner spectroscopic redshift, \(z_\mathrm{F} = {\bar z_{\mathrm{spec}}}\), as described in Section~\ref{subsection:discovery.galaxy_template}. Column 4 provides the spectroscopically detected background emission-line source redshift, \(z_\mathrm{B}\), computed as described in Section~\ref{subsection:discovery.inspection}.  Column 5 provides the NSA Sersic effective radius. Column 6 provides the Einstein radius with upper and lower limits. Column 7 provides the number of good background emission-line hits identified within the common source-plane. For those source-planes with only a single hit, we distinguish the case from one multi-line hit (identified by a $\dagger$) from the case of one single-line hit (identified by *).  Column 8 provides the ratio of the radial distance of a source-plane image's detectable edge to the estimated upper limit of the Einstein radius.}
\centering
\begin{tabular}{ l c c c c c c c c c c }
\hline
\hline
{System Name} & {Plate--IFU} & {\(z_{\mathrm{F}}\)} & {\(z_{\mathrm{B}}\)} & {\(R_{\mathrm{eff}}~[\arcsec]\)} & {\(R_{\mathrm{Ein}}~[\arcsec]\)} & {Hit No.} & {\(R/R_\mathrm{Ein}^{\mathrm{Upper}}\)} & \\
{\scriptsize (1)} & {\scriptsize (2)} & {\scriptsize (3)} & {\scriptsize (4)} & {\scriptsize (5)} & {\scriptsize (6)} & {\scriptsize (7)} & {\scriptsize (8)} \\
\hline

SDSS~J034957.70+000427.60 & 8086--6101 & 0.13155 & 0.669 & 5.13 & $1.20_{-1.02}^{+0.40}$ & 3 & 0.80 \\
SDSS~J035725.22-052514.40 & 8728--3704 & 0.06405 & 0.202 & 7.54 & $2.32_{-1.59}^{+0.29}$ & 20 & 0.73 \\
SDSS~J072816.98+400502.78 & 8131--6102 & 0.04956 & 0.694 & 12.34 & $2.74_{-1.56}^{+0.13}$ & 1$^{\dagger}$ & 1.12 \\
SDSS~J072816.98+400502.78 & 8131--6102 & 0.04956 & 0.954 & 12.34 & $2.77_{-1.57}^{+0.19}$ & 1$^{*}$ & 0.67 \\
SDSS~J073325.27+441111.32$^{\dagger}$ & 8133--6101 & 0.05558 & 0.778 & 5.4 & $0.23_{-0.06}^{+0.00}$ & 1$^{*}$ & 9.70 \\
SDSS~J074752.22+432300.60 & 8137--6104 & 0.11144 & 0.863 & 11.65 & $1.13_{-0.98}^{+0.06}$ & 1$^{\dagger}$ & 2.39 \\
SDSS~J075106.24+472000.07 & 8714--6103 & 0.07885 & 0.778 & 10.65 & $2.13_{-1.38}^{+0.21}$ & 1$^{\dagger}$ & 0.15 \\
SDSS~J082117.33+252930.73 & 8939--6103 & 0.09487 & 1.165 & 8.13 & $2.09_{-1.35}^{+0.25}$ & 2 & 0.77 \\
SDSS~J082519.86+172855.36$^{\dagger}$ & 8241--3701 & 0.08898 & 0.800 & 4.64 & $1.20_{-1.04}^{+0.43}$ & 2 & 4.13 \\
SDSS~J083633.11+540229.27 & 8243--12703 & 0.03052 & 0.482 & 25.68 & $0.76_{-0.62}^{+0.37}$ & 60 & 0.88 \\
SDSS~J084531.60+535732.41 & 8243--3704 & 0.03083 & 0.228 & 4.76 & $0.73_{-0.56}^{+0.03}$ & 51 & 1.03 \\
SDSS~J090138.81+274002.83 & 8987--12701 & 0.14375 & 0.388 & 11.77 & $0.20_{-0.06}^{+0.32}$ & 24 & 2.25 \\
SDSS~J102726.54+375237.35 & 8568--9102 & 0.09802 & 0.304 & 7.13 & $0.74_{-0.55}^{+0.40}$ & 1$^{\dagger}$ & 1.80 \\
SDSS~J105953.06+500054.72 & 8948--12702 & 0.02571 & 1.100 & 24.74 & $0.75_{-0.53}^{+0.38}$ & 2 & 0.49 \\
SDSS~J110602.37+431024.50 & 8255--6101 & 0.05837 & 0.789 & 5.55 & $0.75_{-0.54}^{+0.40}$ & 4 & 0.44 \\
SDSS~J110814.21+433729.34 & 8255--6104 & 0.07561 & 0.382 & 10.52 & $1.88_{-1.36}^{+0.26}$ & 31 & 1.50 \\
SDSS~J112034.12+464928.22$^{\dagger}$ & 8946--9102 & 0.05355 & 0.713 & 13.82 & $2.60_{-1.86}^{+0.16}$ & 1$^{*}$ & 0.69 \\
SDSS~J112312.82+511109.19$^{\dagger}$ & 8947--6104 & 0.04865 & 0.165 & 10.22 & $1.85_{-1.13}^{+0.06}$ & 2 & 1.91 \\
SDSS~J112312.82+511109.19$^{\dagger}$ & 8947--6104 & 0.04865 & 0.264 & 10.22 & $2.27_{-1.50}^{+0.08}$ & 22 & 2.13 \\
SDSS~J112907.67+213509.63 & 8450--12704 & 0.04557 & 0.449 & 12.37 & $2.57_{-1.81}^{+0.16}$ & 3 & 1.18 \\
SDSS~J130701.21+461833.95$^{\dagger}$ & 8318--1901 & 0.02451 & 0.141 & 3.83 & $0.18_{-0.04}^{+0.00}$ & 3 & 6.09 \\
SDSS~J131826.10+471323.40 & 8319--12701 & 0.05711 & 0.222 & 13.21 & $1.23_{-0.99}^{+0.38}$ & 28 & 0.85 \\
SDSS~J143607.49+494313.22$^{\dagger}$ & 8595--6101 & 0.12538 & 1.231 & 3.69 & $2.59_{-1.84}^{+0.35}$ & 2 & 0.63 \\
SDSS~J144835.21+512308.07 & 8592--12705 & 0.13151 & 0.433 & 8.66 & $0.79_{-0.62}^{+0.38}$ & 21 & 1.27 \\
SDSS~J153101.12+285253.59 & 9042--3701 & 0.08275 & 0.751 & 9.29 & $2.79_{-1.19}^{+0.14}$ & 2 & 1.76 \\
SDSS~J154014.39+483529.51 & 8485--6104 & 0.06582 & 0.487 & 4.75 & $0.75_{-0.57}^{+0.41}$ & 17 & 1.11 \\
SDSS~J154307.00+390904.85 & 8315--6104 & 0.06382 & 0.258 & 11.35 & $1.25_{-1.01}^{+0.39}$ & 4 & 2.07 \\
SDSS~J161711.62+495751.83 & 8482--12701 & 0.05739 & 0.826 & 15.78 & $1.86_{-1.34}^{+0.40}$ & 1$^{*}$ & 0.82 \\
SDSS~J162332.75+390715.95 & 8604--12701 & 0.03505 & 0.250 & 23.61 & $0.75_{-0.57}^{+0.41}$ & 4 & 0.79 \\
SDSS~J162526.09+395214.51$^{\dagger}$ & 8550--12701 & 0.02907 & 0.994 & 14.8 & $1.15_{-0.97}^{+0.06}$ & 3 & 1.71 \\
SDSS~J162737.71+423817.63 & 9029--3703 & 0.03107 & 0.709 & 7.58 & $1.70_{-1.18}^{+0.23}$ & 2 & 1.92 \\
SDSS~J162920.74+394913.43 & 8602--1901 & 0.02608 & 0.923 & 4.83 & $1.23_{-1.04}^{+0.04}$ & 2 & 1.23 \\
SDSS~J164028.01+391912.42 & 8588--6102 & 0.03005 & 0.962 & 10.76 & $2.58_{-2.06}^{+0.18}$ & 3 & 1.09 \\
SDSS~J164943.29+385221.78 & 8612--6103 & 0.06150 & 0.290 & 12.58 & $2.37_{-1.22}^{+0.25}$ & 25 & 1.83 \\
SDSS~J170007.17+375022.21 & 8606--12701 & 0.06330 & 0.797 & 25.2 & $1.19_{-0.96}^{+0.04}$ & 6 & 0.54 \\
SDSS~J170124.01+372258.09$^{\dagger}$ & 8606--6102 & 0.12167 & 0.790 & 11.15 & $1.91_{-1.18}^{+0.34}$ & 30 & 0.37 \\
SDSS~J173042.33+563821.79 & 8626--3701 & 0.02938 & 0.797 & 6.68 & $0.73_{-0.55}^{+0.02}$ & 1$^{*}$ & 0.78 \\
SDSS~J221101.52+122227.49 & 7977--6102 & 0.06139 & 0.775 & 12.06 & $1.86_{-1.34}^{+0.28}$ & 1$^{*}$ & 1.80 \\
\hline
\end{tabular}
\end{table*}

The following selection criteria were applied to select the strong-lensing candidates from the source-planes:
\begin{itemize}
\item The source-plane contained at least one multi-line hit with a minimum of three emission-lines, or
\item Each source-plane contained at least two (single-line or multi-line) hits, or
\item The source-plane contained at least one (single-line or multi-line) hit with additional evidence of low S/N emission-lines that are correlated with the source-plane redshift, and are contained within fibres that are positioned near the hit.
\item The source-plane contained at least one background emission-line detected at a position within twice the upper limit of the Einstein radius.
\end{itemize}

We made an exception for system SDSS J0733+4411, SDSS J0825+1728, and SDSS J1307+4618, since each displayed well formed lensing features (see Section~\ref{section:lensing}). After applying these selections, we found 36 promising candidates out of which 2 showed evidence for compound lensing with multiple source-planes (see Section~\ref{section:compound}). This final sample is presented in Table~\ref{table:source_inspection}.

Figure~\ref{figure.hits2d} provides two-dimensional plots of each of these 36 candidate lenses, showing each of the above selection criteria including the positions of the high signal-to-noise emission-line hits, and the location of the manually confirmed evidence of low S/N emission-lines from the background source. The plots also show the upper limit of the Einstein radii and the colour-scale of the square markers indicates the gravitational lensing convergence, computed as 
\begin{equation}\label{equation.kappa}
\kappa = \Sigma/\Sigma_{cr},
\end{equation}
in order to indicate the strong lensing regime where \(\kappa \gtrsim 1\), where \(\Sigma\) is the surface-mass density determined from the MaNGA Firefly VAC discussed in Section~\ref{subsection:discovery.einstein}, and 
\begin{equation}\label{equation.sigmacrit}
\Sigma_{cr} = \frac{c^2}{4 \pi G} \frac{D_{\mathrm S}}{D_{\mathrm L} D_{\mathrm LS}}
\end{equation}
is the critical surface-mass density.  

\newcommand{\hitscaption}{Plots of background emission line hits for each SILO candidate. The positions of the hits are shown with cyan numbers representing the good hit count, and the cyan triangles show the location of the manually confirmed evidence of low S/N background emission-lines. The inner and outer radii of the cyan circle represent the lower and upper limits of the Einstein radius. The green coloured circle represents twice the upper limit of the Einstein radius, and thus the upper limit of the strong lensing regime.  The value of the gravitational lensing convergence, \(\kappa = \Sigma/\Sigma_{cr}\), according to Equation~\ref{equation.kappa}, is plotted with squares with respect to the colour-bar scale. An image of the foreground galaxy is underlaid for the purpose of visual orientation, including a $5\arcsec$ ruler at the top-left. The pink hexagon displays the field of view within the MaNGA IFU.}

\begin{figure*}
\begin{center}
\label{figure.hits2d}
\includegraphics[page=1, width=1\textwidth]{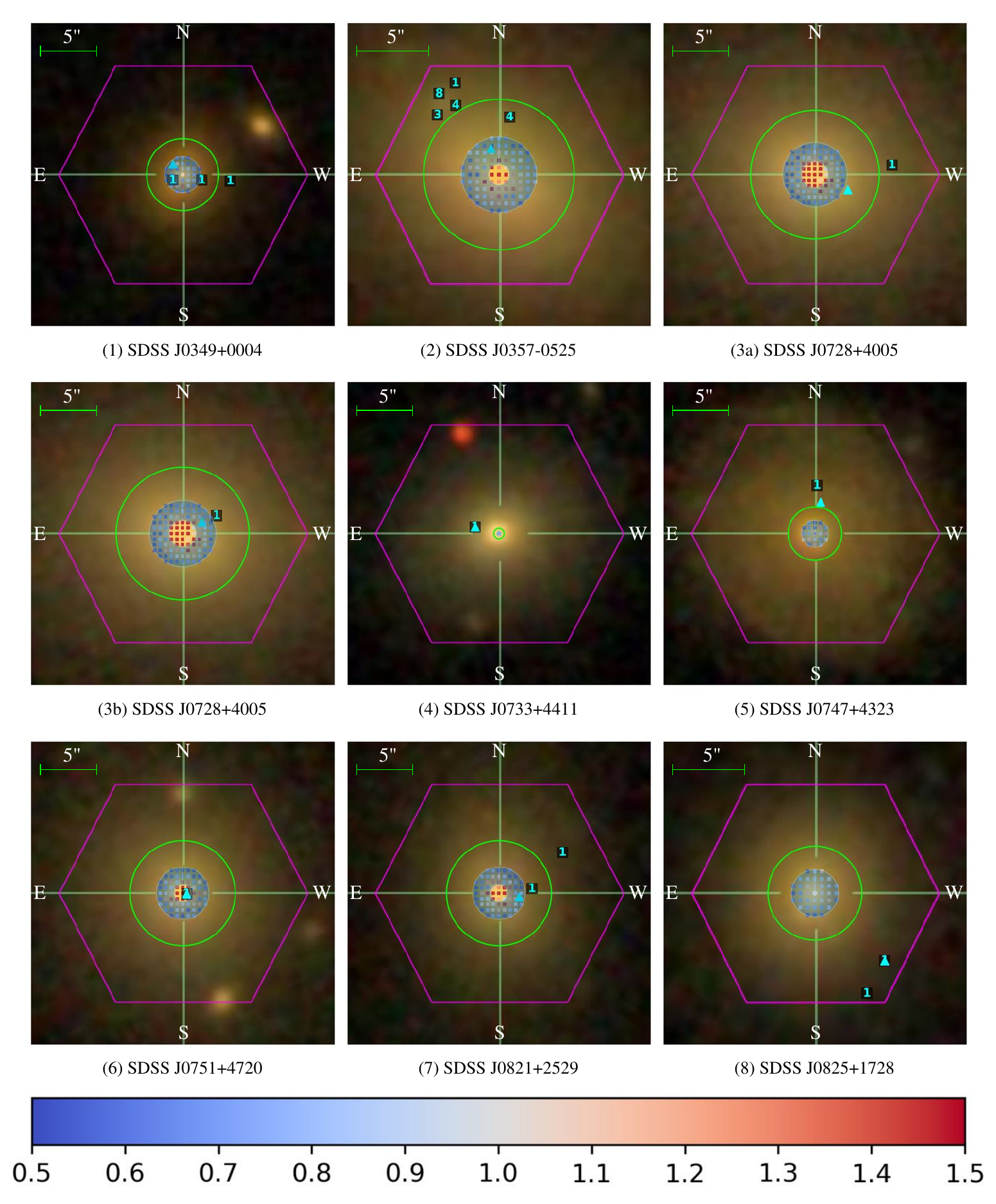}
\end{center}
\caption{\hitscaption\ This figure is continued.}
\end{figure*}
\addtocounter{figure}{-1}

\begin{figure*}
\begin{center}
\includegraphics[page=1, width=1\textwidth]{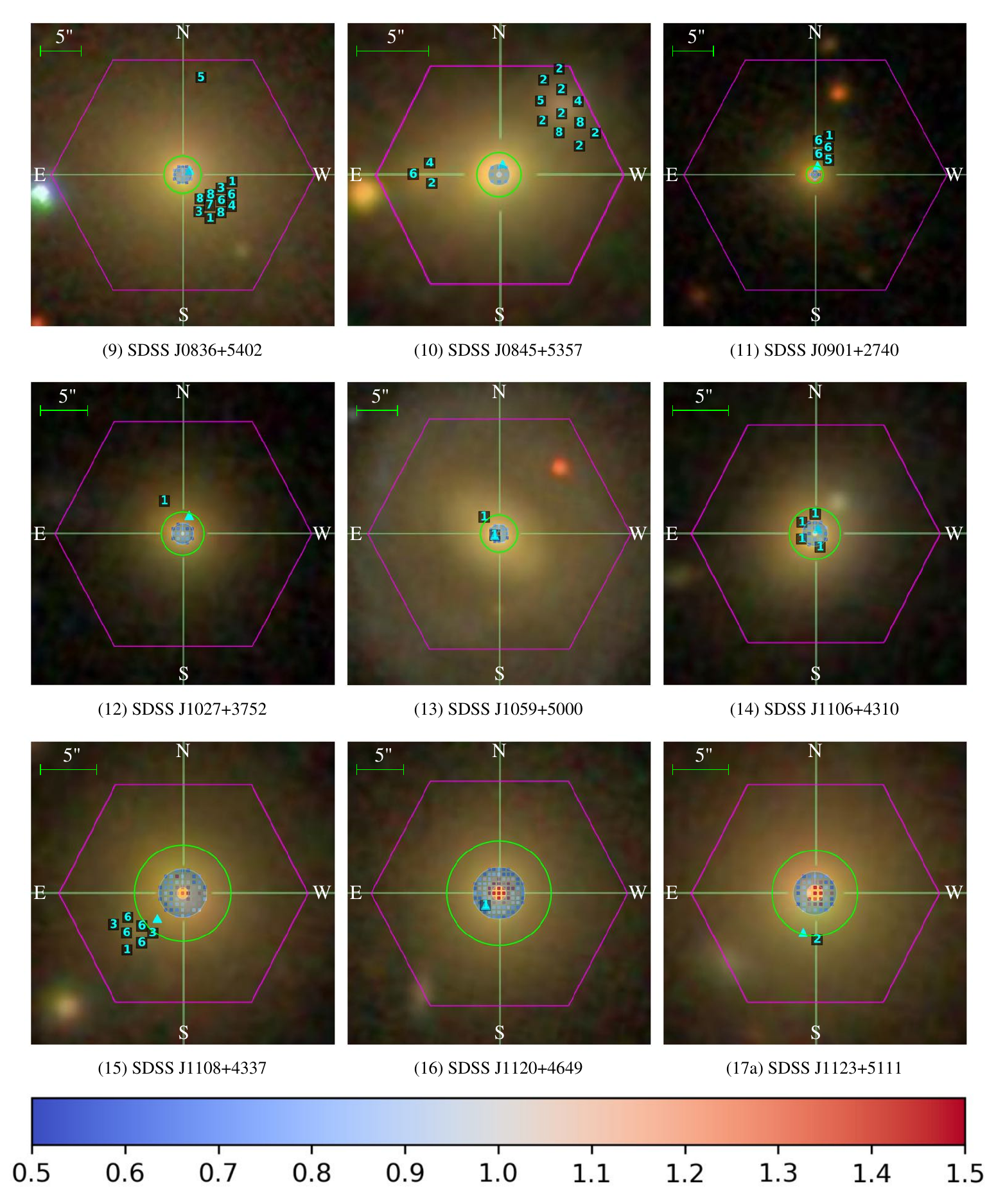}
\end{center}
\caption{\emph{Continued.} \hitscaption\  This figure is continued.}
\end{figure*}
\addtocounter{figure}{-1}

\begin{figure*}
\begin{center}
\includegraphics[page=1, width=1\textwidth]{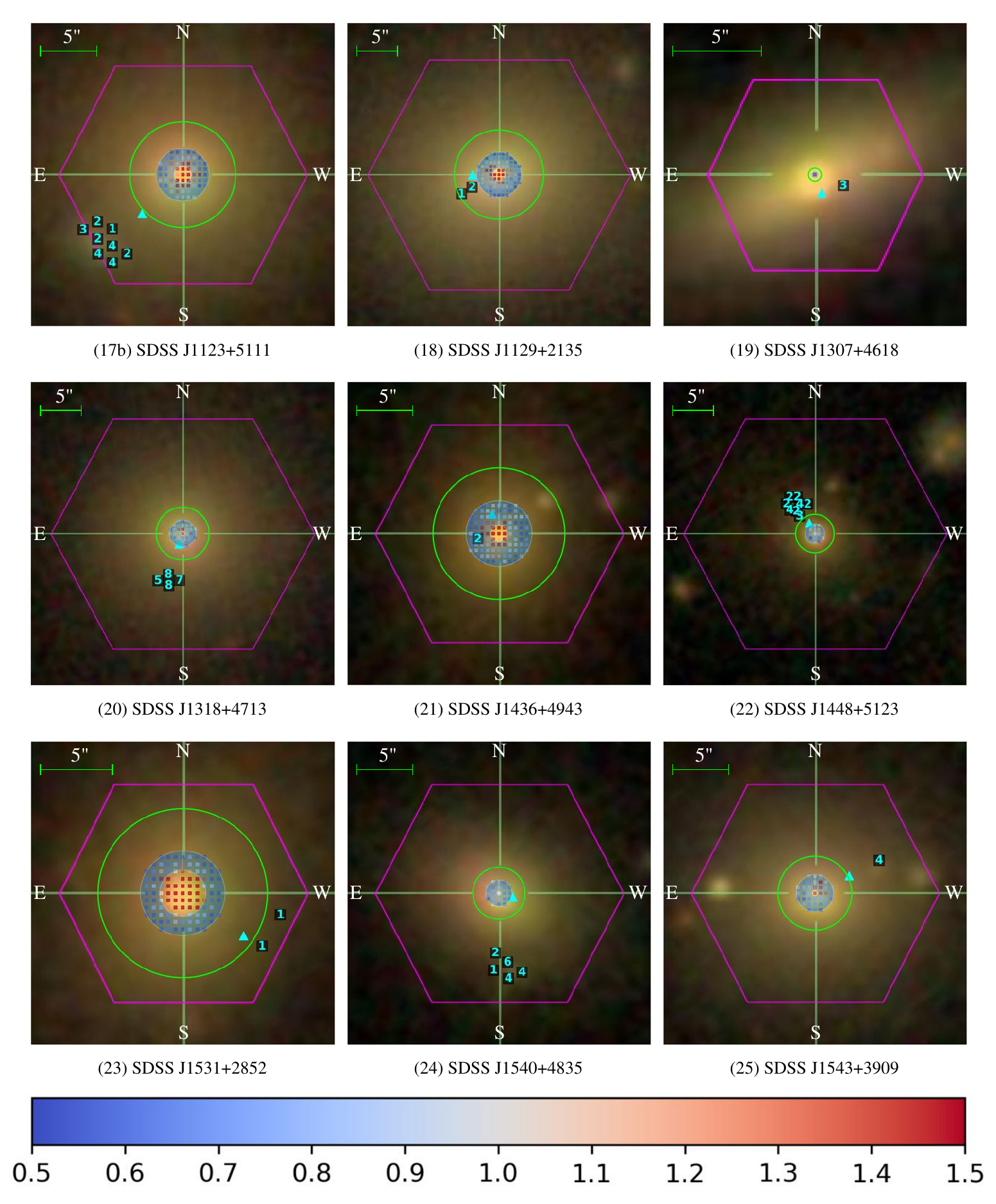}
\end{center}
\caption{\emph{Continued.} \hitscaption\  This figure is continued.}
\end{figure*}
\addtocounter{figure}{-1}

\begin{figure*}
\begin{center}
\includegraphics[page=1, width=1\textwidth]{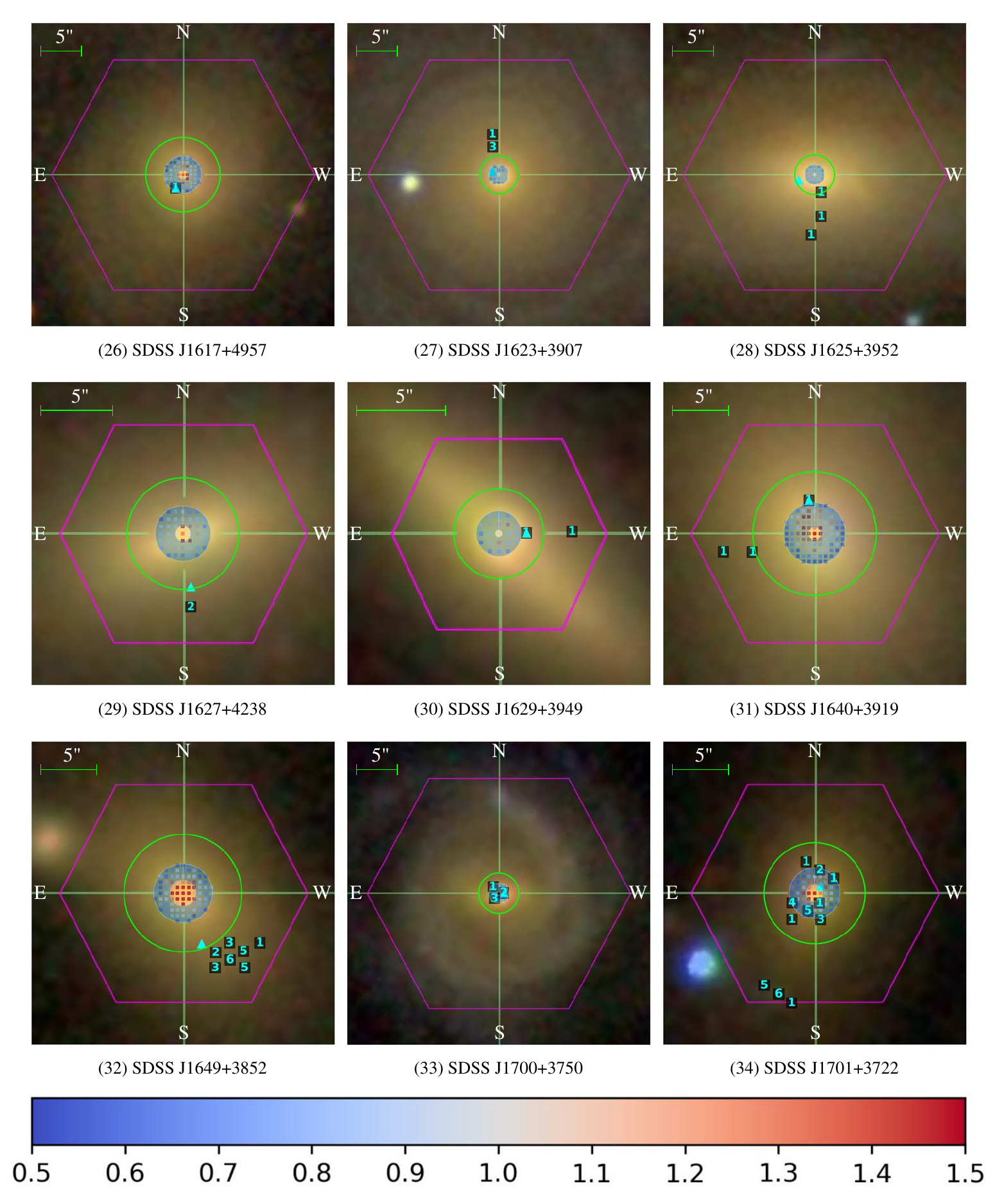}
\end{center}
\caption{\emph{Continued.} \hitscaption\  This figure is continued.}
\end{figure*}
\addtocounter{figure}{-1}

\begin{figure*}
\begin{center}
\includegraphics[page=1, width=1\textwidth]{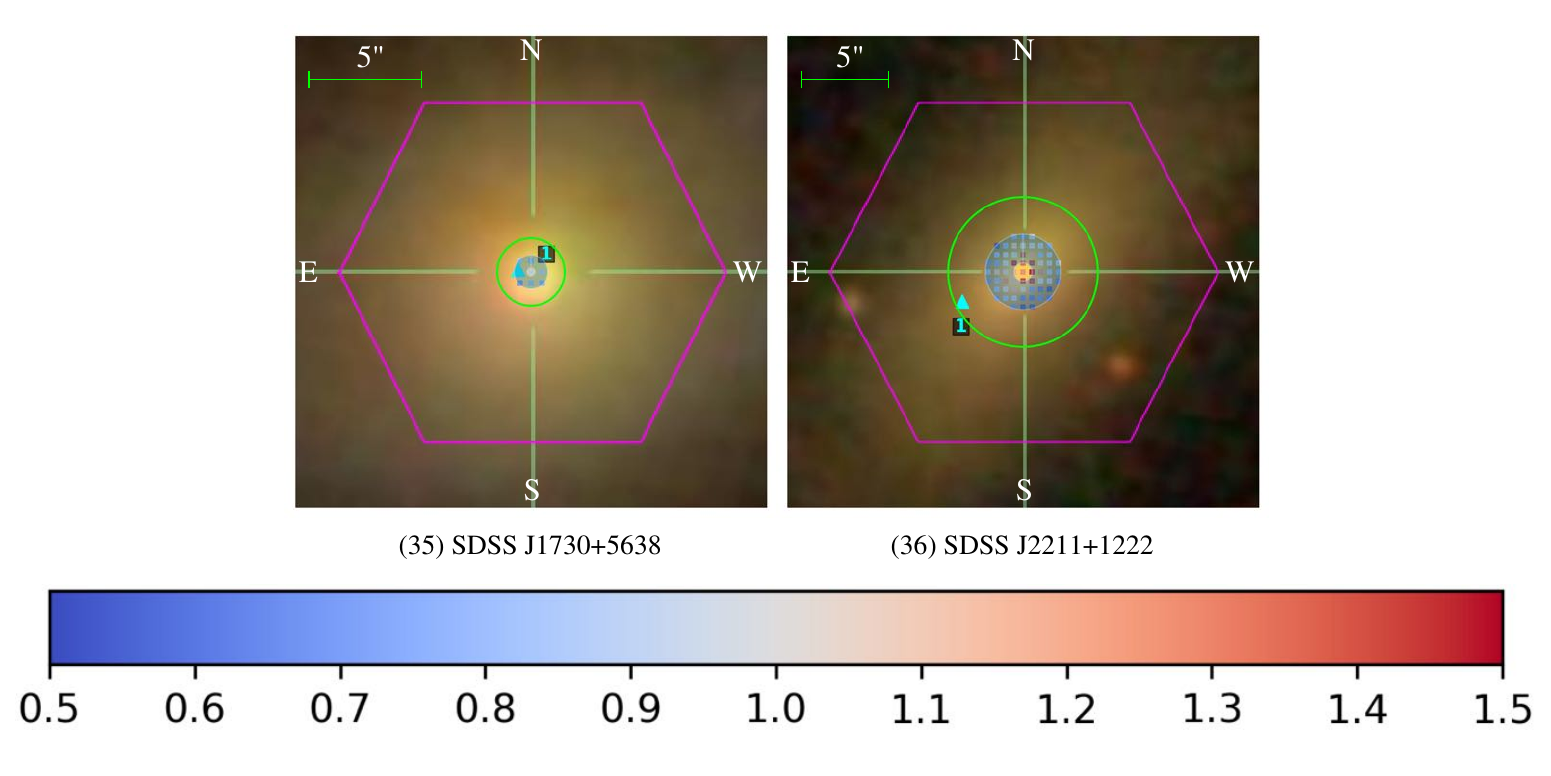}
\end{center}
\caption{\emph{Continued.} \hitscaption}
\end{figure*}

\section{Narrow-band Lensed Image Construction}\label{section:lensing}

Since the candidates have a spectroscopically determined background redshift, provided by \(z_\mathrm{B}\) as listed in Table~\ref{table:emline}, we are able to construct narrow-band images from the spectra.  Although we do not generally have sufficient spatial resolution to identify gravitational lensing counter-images in all of the narrow-band images, we did serendipitously find some examples based on the following technique: 

Using the background spectroscopic redshifts in Section~\ref{subsection:discovery.emline}, we are able to construct narrow-band images around specific background emission-lines. In order to obtain smooth maps, we used the MaNGA DRP Cube files, which provided interpolated spectra for each spaxel~\citep{2016AJ....152...83L}. To subtract the flux of the foreground galaxy, we fitted a model for each spaxel using the same method described in Section~\ref{subsection:discovery.galaxy_template}. We summed each spaxel's residual flux within 4--8~\AA\ within the wavelengths of the [O\,{\sc ii}], [O\,{\sc iii}a], and H$\alpha$ background emission-lines. The two-dimensional maps of these combined narrow-band background emission-lines are plotted in Figure~\ref{figure:narrowband}.

\newcommand{\narrowbandcaption}{Images of galaxies that show strong lensing features in narrow-band images.  The left panel shows the foreground galaxy (MaNGA target) produced by the MaNGA Data Reduction Pipeline~\citep[DRP;][]{2016AJ....152...83L}.  The pink hexagon displays the field of view within the target IFU. The right panel shows the narrow-band image created by combining the flux of the [O\,{\sc ii}], [O\,{\sc iii}a], and H$\alpha$ emission-lines within a \(\pm 4\)\AA\ -- \(\pm 8\)\AA\ window at the redshift of the background source-plane, as described in Section~\ref{section:lensing}.  High signal to noise detections are marked with cyan circles, and are likely images of background galaxies. Spaxels that show low S/N images or counter-images are marked with magenta crosses, and add support of strong lensing because they are in the same source-plane.  Each panel displays a $5\arcsec$ ruler at the top-left.  The foreground and background redshifts are shown at the bottom-left and bottom-right, respectively.  Panel (4) SDSS~J1123+5111 is a possible compound lens, with two source-planes.}
\begin{figure*}
\begin{center}
\includegraphics[width=1\textwidth]{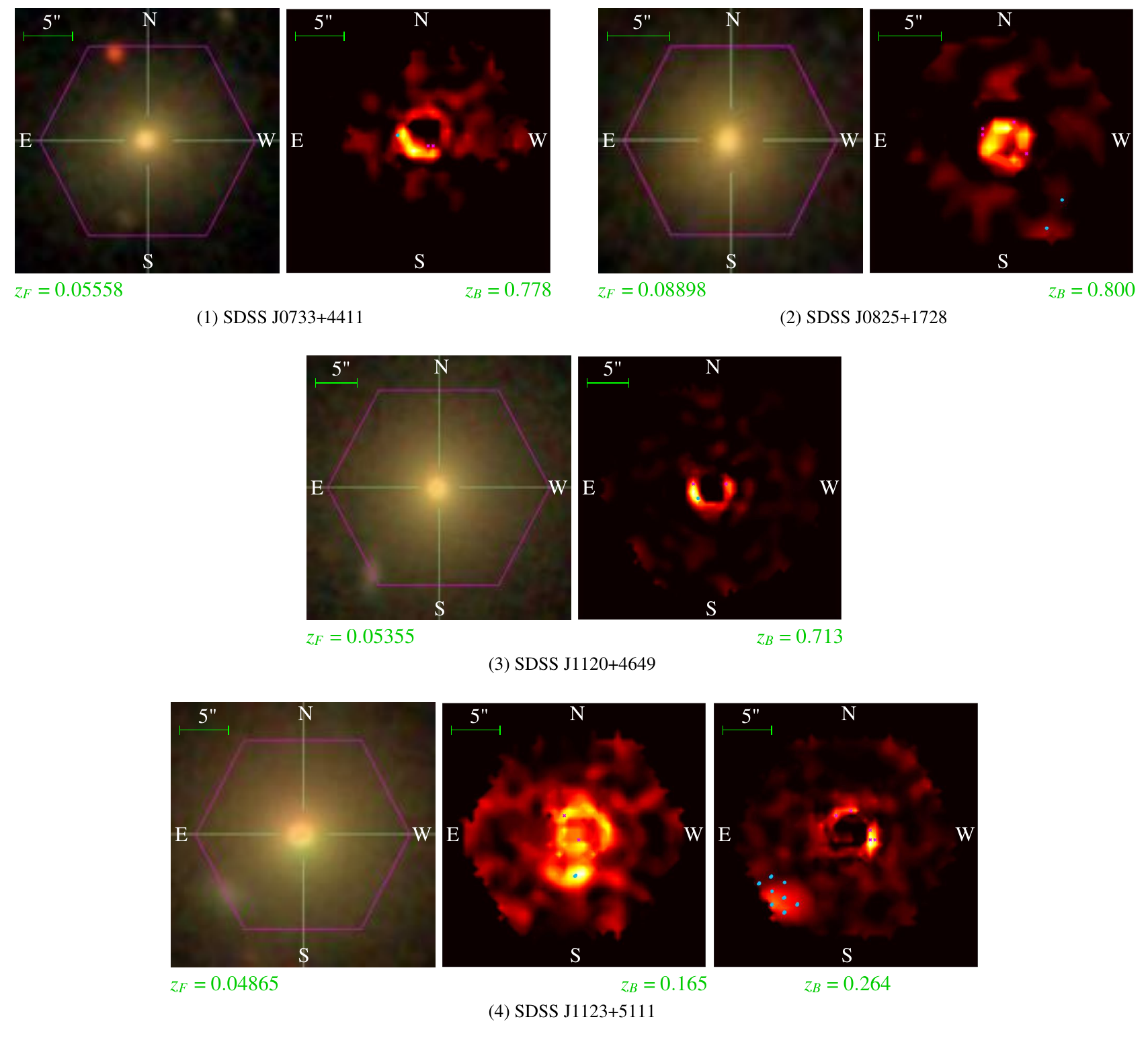}
\end{center}
\caption{\narrowbandcaption}
\label{figure:narrowband}
\end{figure*}

\addtocounter{figure}{-1}
\begin{figure*}
\begin{center}
\includegraphics[width=1\textwidth]{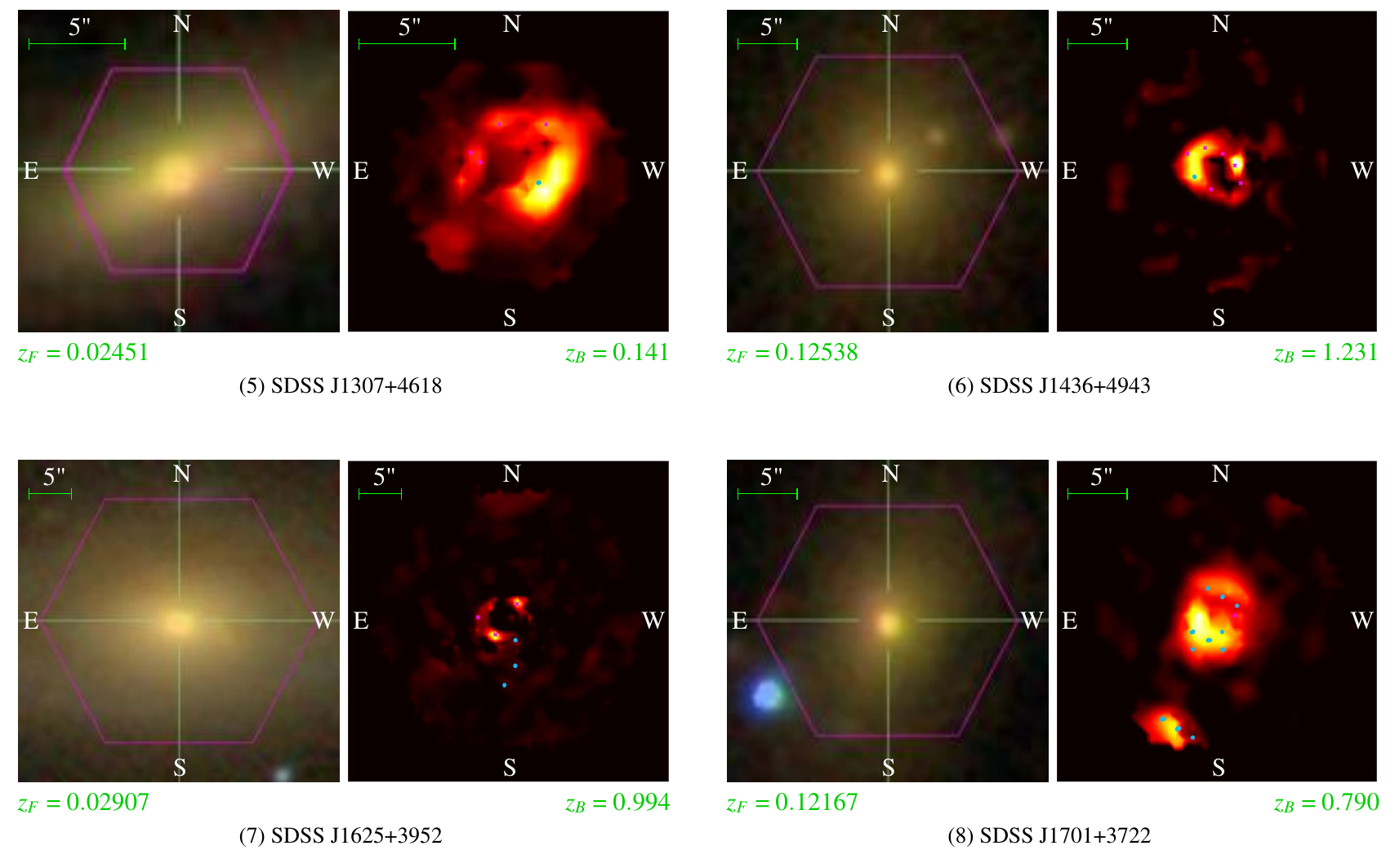}
\end{center}
\caption{\narrowbandcaption}
\end{figure*}

For most galaxies, the counter-image is near the foreground galaxy core. Although we have subtracted the foreground flux from the narrow-band images, the error in the residual flux is larger near the centre where the foreground is brightest whereas the counter-image is faint. Therefore, the chances to observe evidence of counter-images is small. Despite the challenges, we discovered 8 candidates with counter-images shown in Figure~\ref{figure:narrowband}.  One of these systems, shown in Panel (4), is a compound lens candidate with 2 background source-planes, as further discussed in Section~\ref{section:compound}.

For each galaxy in Figure~\ref{figure:narrowband} we indicate the positions (with cyan circles) of the high S/N detected background emission-lines that were visually inspected, as described in Section~\ref{subsection:discovery.inspection}. We manually inspected the source-plane for emission-lines that were otherwise too weak to be detected in the initial search, shown with magenta crosses in each spaxel.

Although there are several sources of uncertainty in the construction of the narrow-band images, and follow-up imaging is required to confirm our interpretation, we observe the following lensing features, as described in Table~\ref{table:narrowband_inspection}.

\begin{table*}
\caption{\label{table:narrowband_inspection}Narrow-band Inspection. This list identifies the strong lensing features observed within the narrow-band images from Figure~\ref{figure:narrowband}. Column 1 presents the SDSS System Name. Column 2 provides the manual inspection which identifies the strong lensing features.}
\centering
\begin{tabular}{ l l }
\hline
\hline
{System Name} & {Manual Inspection} \\
\phantom{Syste }{\scriptsize (1)} & \phantom{Manual }{\scriptsize (2)} \\
\hline
SDSS~J0733+4411 & {Possible ring.} \\
SDSS~J0825+1728 & {Possible quad.} \\
SDSS~J1120+4649 & {Image with a isolated counter-image.} \\
SDSS~J1123+5111 & \parbox{14.5cm}{Candidate compound lens with two background source planes: Image with a visible counter-image of an extended source, and a higher redshift image with a visible counter-image.} \\
SDSS~J1307+4618 & {Multiple images.} \\
SDSS~J1436+4943 & {Image with a isolated counter-image.} \\
SDSS~J1625+3952 & {Image with a isolated counter-image of high redshift background galaxy.} \\
SDSS~J1701+3722 & {Image with a isolated counter-image, confirming \cite{2017MNRAS.464L..46S}.} \\
\hline
\end{tabular}
\end{table*}
\section{Candidate multiple source-plane lenses}\label{section:compound}

Two of our candidate lenses show high signal-to-noise background emission-lines from distinct background source planes:
\begin{itemize}
\item {\bf SDSS~J0728+4005} listed in Table~\ref{table:source_inspection} with foreground redshift \(z_{\mathrm F}=0.04956\), and shown in Panels 3a and 3b of Figure~\ref{figure.hits2d}, with background redshifts \(z_{\mathrm B}=0.694\) and \(z_{\mathrm B}=0.954\),
\item {\bf SDSS~J1123+5111} listed in Table~\ref{table:source_inspection} with foreground redshift \(z_{\mathrm F}=0.04865\), and shown in Panels 17a and 17b of Figure~\ref{figure.hits2d} and Panel 4 of Figure~\ref{figure:narrowband}, with background redshifts \(z_{\mathrm B}=0.165\) and \(z_{\mathrm B}=0.264\).
\end{itemize}

The discovery of such systems is very rare but also of great scientific value due to having multiple Einstein radii to probe the mass distribution at distinct positions within the halo.  Since the Einstein radius is a function of the angular diameter distance and the mass enclosed within the Einstein radius, these systems provide novel constraints on the total mass density profile of the lens~\citep{2012ApJ...752..163S, 2014MNRAS.443..969C}. The ratio of Einstein radii in multiple source-plane lenses is sensitive to cosmological parameters, but independent of the Hubble constant~\citep{2012MNRAS.424.2864C}. Our candidates do not have optimal redshift configurations for constraining dark energy models with a constant equation of state, but they are potentially exciting probes of more complicated models of dark energy since low redshift compound lenses have a unique sensitivity to variations in the dark energy equation of state \citep{2017APh....86...41L}.

\section{Summary and Conclusions}\label{section:conclusions}

We introduced the Spectroscopic Identification of Lensing Objects (SILO) survey, and presented a catalogue of 38 strong gravitational lens candidates discovered within the 2812 MaNGA galaxies released with SDSS-IV DR14, of which 8 have Grade-A lensing features visible in narrow-band images of the background source-plane. Out of the $\sim1.5$ million MaNGA fibres, $698$ exhibit background emission-lines above a threshold signal-to-noise ratio allowing a partial reconstruction of the source-plane geometry. Two of these systems are compound lens candidates, one of which demonstrates lensing features visible within the narrow-band image.

These SILO candidates complement previous SLACS, SWELLS, S4TM, BELLS, and BELLS GALLERY lenses by exploring the population of galaxies at smaller distances and larger Einstein radius. Follow-up imaging of these candidates could potentially confirm our interpretation of the detected background emission-lines within our estimated Einstein radii as strongly lensed images of more distant galaxies.

Combined with the stellar density maps, the lens models we obtain from follow-up imaging would generate mass profile parameters at positions with smaller \(R/R_\mathrm{eff}\) than currently available. These additional constraints at small scales can be used to break the degeneracy between luminous and dark matter that currently plagues dynamical studies, and shed light on the bulge-halo conspiracy \citep{2006ApJ...640..662T, 2014MNRAS.438.3594D, 2017ApJ...840...34S}. Lenses with a low \(R/R_\mathrm{eff}\) ratio are particularly useful to obtain estimates (and possibly gradients) of the stellar initial mass function in the inner regions of massive galaxies, where the mass budget is dominated by the baryons \citep[see e.g.][]{2017ApJ...845..157N, 2017ApJ...844L..11V}.

Because the SILO survey extends the spectroscopic discovery method of \cite{2006ApJ...638..703B, 2012ApJ...744...41B} to multiple IFU spectra, a follow-up imaging survey would provide validation of the source-plane distribution of lensing features and narrow-band images which were used in the manual inspection process.  This would allow us to refine our catalogue grading procedure. A further extension of the spectroscopic discovery method to include high-redshift Lyman-$\alpha$ emitter source galaxies, following the procedure of the BELLS GALLERY survey \citep{2016ApJ...824...86S}, can be expected to increase the number of candidate lenses.  Since MaNGA will observe a total of 10,000 galaxies during SDSS-IV, we expect a large database of background emission-lines, from which a catalogue of confirmed strong gravitational lenses would be constructed, which would ultimately be of great significance to many future studies. 
\section*{Acknowledgements}

Funding for the Sloan Digital Sky Survey IV has been provided by the Alfred P. Sloan Foundation, the U.S. Department of Energy Office of Science, and the Participating Institutions. SDSS-IV acknowledges support and resources from the Center for High-Performance Computing at the University of Utah. The SDSS web site is \href{http://www.sdss.org}{www.sdss.org}.

SDSS-IV is managed by the Astrophysical Research Consortium for the  Participating Institutions of the SDSS Collaboration including the  Brazilian Participation Group, the Carnegie Institution for Science,  Carnegie Mellon University, the Chilean Participation Group, the French Participation Group, Harvard-Smithsonian Center for Astrophysics,  Instituto de Astrof\'isica de Canarias, The Johns Hopkins University,  Kavli Institute for the Physics and Mathematics of the Universe (IPMU) /  University of Tokyo, Lawrence Berkeley National Laboratory,  Leibniz Institut f\"ur Astrophysik Potsdam (AIP),   Max-Planck-Institut f\"ur Astronomie (MPIA Heidelberg),  Max-Planck-Institut f\"ur Astrophysik (MPA Garching),  Max-Planck-Institut f\"ur Extraterrestrische Physik (MPE),  National Astronomical Observatories of China, New Mexico State University,  New York University, University of Notre Dame,  Observat\'ario Nacional / MCTI, The Ohio State University,  Pennsylvania State University, Shanghai Astronomical Observatory,  United Kingdom Participation Group, Universidad Nacional Aut\'onoma de M\'exico, University of Arizona,  University of Colorado Boulder, University of Oxford, University of Portsmouth,  University of Utah, University of Virginia, University of Washington, University of Wisconsin,  Vanderbilt University, and Yale University.

This work was supported by World Premier International Research Center Initiative (WPI Initiative), MEXT, Japan. AS is partly supported by KAKENHI Grant Number JP17K14250. AW acknowledges support of a Leverhulme Trust Early Career Fellowship.

This research made use of {\sc marvin}~\citep{zenodo.292632}, a core Python package and web framework for MaNGA data, developed by Brian Cherinka, Jos\'e S\'anchez-Gallego, and Brett H.\,Andrews, and Joel R.\,Brownstein. (SDSS-IV MaNGA Collaboration, 2018).

\bibliographystyle{mnras}
\bibliography{tex/references}

\begin{thebibliography}{}
\makeatletter
\relax
\def\mn@urlcharsother{\let\do\@makeother \do\$\do\&\do\#\do\^\do\_\do\%\do\~}
\def\mn@doi{\begingroup\mn@urlcharsother \@ifnextchar [ {\mn@doi@}
  {\mn@doi@[]}}
\def\mn@doi@[#1]#2{\def\@tempa{#1}\ifx\@tempa\@empty \href
  {http://dx.doi.org/#2} {doi:#2}\else \href {http://dx.doi.org/#2} {#1}\fi
  \endgroup}
\def\mn@eprint#1#2{\mn@eprint@#1:#2::\@nil}
\def\mn@eprint@arXiv#1{\href {http://arxiv.org/abs/#1} {{\tt arXiv:#1}}}
\def\mn@eprint@dblp#1{\href {http://dblp.uni-trier.de/rec/bibtex/#1.xml}
  {dblp:#1}}
\def\mn@eprint@#1:#2:#3:#4\@nil{\def\@tempa {#1}\def\@tempb {#2}\def\@tempc
  {#3}\ifx \@tempc \@empty \let \@tempc \@tempb \let \@tempb \@tempa \fi \ifx
  \@tempb \@empty \def\@tempb {arXiv}\fi \@ifundefined
  {mn@eprint@\@tempb}{\@tempb:\@tempc}{\expandafter \expandafter \csname
  mn@eprint@\@tempb\endcsname \expandafter{\@tempc}}}

\bibitem[\protect\citeauthoryear{{Albareti} et~al.,}{{Albareti}
  et~al.}{2017}]{2017ApJS..233...25A}
{Albareti} F.~D.,  et~al., 2017, \mn@doi [\apjs] {10.3847/1538-4365/aa8992},
  \href {http://adsabs.harvard.edu/abs/2017ApJS..233...25A} {233, 25}

\bibitem[\protect\citeauthoryear{{Arneson}, {Brownstein}  \&
  {Bolton}}{{Arneson} et~al.}{2012}]{2012ApJ...753....4A}
{Arneson} R.~A.,  {Brownstein} J.~R.,   {Bolton} A.~S.,  2012, \mn@doi [\apj]
  {10.1088/0004-637X/753/1/4}, \href
  {http://adsabs.harvard.edu/abs/2012ApJ...753....4A} {753, 4}

\bibitem[\protect\citeauthoryear{{Auger}, {Treu}, {Bolton}, {Gavazzi},
  {Koopmans}, {Marshall}, {Bundy}  \& {Moustakas}}{{Auger}
  et~al.}{2009}]{2009ApJ...705.1099A}
{Auger} M.~W.,  {Treu} T.,  {Bolton} A.~S.,  {Gavazzi} R.,  {Koopmans}
  L.~V.~E.,  {Marshall} P.~J.,  {Bundy} K.,   {Moustakas} L.~A.,  2009, \mn@doi
  [\apj] {10.1088/0004-637X/705/2/1099}, \href
  {http://adsabs.harvard.edu/abs/2009ApJ...705.1099A} {705, 1099}

\bibitem[\protect\citeauthoryear{{Auger}, {Treu}, {Bolton}, {Gavazzi},
  {Koopmans}, {Marshall}, {Moustakas}  \& {Burles}}{{Auger}
  et~al.}{2010}]{2010ApJ...724..511A}
{Auger} M.~W.,  {Treu} T.,  {Bolton} A.~S.,  {Gavazzi} R.,  {Koopmans}
  L.~V.~E.,  {Marshall} P.~J.,  {Moustakas} L.~A.,   {Burles} S.,  2010,
  \mn@doi [\apj] {10.1088/0004-637X/724/1/511}, \href
  {http://adsabs.harvard.edu/abs/2010ApJ...724..511A} {724, 511}

\bibitem[\protect\citeauthoryear{{Barnab{\`e}}, {Czoske}, {Koopmans}, {Treu},
  {Bolton}  \& {Gavazzi}}{{Barnab{\`e}} et~al.}{2009}]{2009MNRAS.399...21B}
{Barnab{\`e}} M.,  {Czoske} O.,  {Koopmans} L.~V.~E.,  {Treu} T.,  {Bolton}
  A.~S.,   {Gavazzi} R.,  2009, \mn@doi [\mnras]
  {10.1111/j.1365-2966.2009.14941.x}, \href
  {http://adsabs.harvard.edu/abs/2009MNRAS.399...21B} {399, 21}

\bibitem[\protect\citeauthoryear{{Barnab{\`e}} et~al.,}{{Barnab{\`e}}
  et~al.}{2012}]{2012MNRAS.423.1073B}
{Barnab{\`e}} M.,  et~al., 2012, \mn@doi [\mnras]
  {10.1111/j.1365-2966.2012.20934.x}, \href
  {http://adsabs.harvard.edu/abs/2012MNRAS.423.1073B} {423, 1073}

\bibitem[\protect\citeauthoryear{{Blanton} et~al.,}{{Blanton}
  et~al.}{2017}]{2017AJ....154...28B}
{Blanton} M.~R.,  et~al., 2017, \mn@doi [\aj] {10.3847/1538-3881/aa7567}, \href
  {http://adsabs.harvard.edu/abs/2017AJ....154...28B} {154, 28}

\bibitem[\protect\citeauthoryear{{Bolton}, {Burles}, {Schlegel}, {Eisenstein}
  \& {Brinkmann}}{{Bolton} et~al.}{2004}]{2004AJ....127.1860B}
{Bolton} A.~S.,  {Burles} S.,  {Schlegel} D.~J.,  {Eisenstein} D.~J.,
  {Brinkmann} J.,  2004, \mn@doi [\aj] {10.1086/382714}, \href
  {http://adsabs.harvard.edu/abs/2004AJ....127.1860B} {127, 1860}

\bibitem[\protect\citeauthoryear{{Bolton}, {Burles}, {Koopmans}, {Treu}  \&
  {Moustakas}}{{Bolton} et~al.}{2006}]{2006ApJ...638..703B}
{Bolton} A.~S.,  {Burles} S.,  {Koopmans} L.~V.~E.,  {Treu} T.,   {Moustakas}
  L.~A.,  2006, \mn@doi [\apj] {10.1086/498884}, \href
  {http://adsabs.harvard.edu/abs/2006ApJ...638..703B} {638, 703}

\bibitem[\protect\citeauthoryear{{Bolton}, {Burles}, {Koopmans}, {Treu},
  {Gavazzi}, {Moustakas}, {Wayth}  \& {Schlegel}}{{Bolton}
  et~al.}{2008a}]{2008ApJ...682..964B}
{Bolton} A.~S.,  {Burles} S.,  {Koopmans} L.~V.~E.,  {Treu} T.,  {Gavazzi} R.,
  {Moustakas} L.~A.,  {Wayth} R.,   {Schlegel} D.~J.,  2008a, \mn@doi [\apj]
  {10.1086/589327}, \href {http://adsabs.harvard.edu/abs/2008ApJ...682..964B}
  {682, 964}

\bibitem[\protect\citeauthoryear{{Bolton}, {Treu}, {Koopmans}, {Gavazzi},
  {Moustakas}, {Burles}, {Schlegel}  \& {Wayth}}{{Bolton}
  et~al.}{2008b}]{2008ApJ...684..248B}
{Bolton} A.~S.,  {Treu} T.,  {Koopmans} L.~V.~E.,  {Gavazzi} R.,  {Moustakas}
  L.~A.,  {Burles} S.,  {Schlegel} D.~J.,   {Wayth} R.,  2008b, \mn@doi [\apj]
  {10.1086/589989}, \href {http://adsabs.harvard.edu/abs/2008ApJ...684..248B}
  {684, 248}

\bibitem[\protect\citeauthoryear{{Bolton} et~al.,}{{Bolton}
  et~al.}{2012}]{2012AJ....144..144B}
{Bolton} A.~S.,  et~al., 2012, \mn@doi [\aj] {10.1088/0004-6256/144/5/144},
  \href {http://adsabs.harvard.edu/abs/2012AJ....144..144B} {144, 144}

\bibitem[\protect\citeauthoryear{{Brownstein} et~al.,}{{Brownstein}
  et~al.}{2012}]{2012ApJ...744...41B}
{Brownstein} J.~R.,  et~al., 2012, \mn@doi [\apj] {10.1088/0004-637X/744/1/41},
  \href {http://adsabs.harvard.edu/abs/2012ApJ...744...41B} {744, 41}

\bibitem[\protect\citeauthoryear{{Bundy} et~al.,}{{Bundy}
  et~al.}{2015}]{2015ApJ...798....7B}
{Bundy} K.,  et~al., 2015, \mn@doi [\apj] {10.1088/0004-637X/798/1/7}, \href
  {http://adsabs.harvard.edu/abs/2015ApJ...798....7B} {798, 7}

\bibitem[\protect\citeauthoryear{{Cherinka}, {S\'anchez-Gallego}, {Andrews}  \&
  {Brownstein}}{{Cherinka} et~al.}{2017}]{zenodo.292632}
{Cherinka} B.,  {S\'anchez-Gallego} J.,  {Andrews} B.~H.,   {Brownstein} J.~R.,
   2017, SDSS Marvin Beta 2.1.0, \mn@doi{10.5281/zenodo.292632}

\bibitem[\protect\citeauthoryear{{Collett} \& {Auger}}{{Collett} \&
  {Auger}}{2014}]{2014MNRAS.443..969C}
{Collett} T.~E.,  {Auger} M.~W.,  2014, \mn@doi [\mnras]
  {10.1093/mnras/stu1190}, \href
  {http://adsabs.harvard.edu/abs/2014MNRAS.443..969C} {443, 969}

\bibitem[\protect\citeauthoryear{{Collett}, {Auger}, {Belokurov}, {Marshall}
  \& {Hall}}{{Collett} et~al.}{2012}]{2012MNRAS.424.2864C}
{Collett} T.~E.,  {Auger} M.~W.,  {Belokurov} V.,  {Marshall} P.~J.,   {Hall}
  A.~C.,  2012, \mn@doi [\mnras] {10.1111/j.1365-2966.2012.21424.x}, \href
  {http://adsabs.harvard.edu/abs/2012MNRAS.424.2864C} {424, 2864}

\bibitem[\protect\citeauthoryear{{Cornachione} et~al.,}{{Cornachione}
  et~al.}{2018}]{2018ApJ...853..148C}
{Cornachione} M.~A.,  et~al., 2018, \mn@doi [\apj] {10.3847/1538-4357/aaa412},
  \href {http://adsabs.harvard.edu/abs/2018ApJ...853..148C} {853, 148}

\bibitem[\protect\citeauthoryear{{Dawson} et~al.,}{{Dawson}
  et~al.}{2013}]{2013AJ....145...10D}
{Dawson} K.~S.,  et~al., 2013, \mn@doi [\aj] {10.1088/0004-6256/145/1/10},
  \href {http://adsabs.harvard.edu/abs/2013AJ....145...10D} {145, 10}

\bibitem[\protect\citeauthoryear{{Despali}, {Vegetti}, {White}, {Giocoli}  \&
  {van den Bosch}}{{Despali} et~al.}{2018}]{2018MNRAS.475.5424D}
{Despali} G.,  {Vegetti} S.,  {White} S.~D.~M.,  {Giocoli} C.,   {van den
  Bosch} F.~C.,  2018, \mn@doi [\mnras] {10.1093/mnras/sty159}, \href
  {http://adsabs.harvard.edu/abs/2018MNRAS.475.5424D} {475, 5424}

\bibitem[\protect\citeauthoryear{{Drory} et~al.,}{{Drory}
  et~al.}{2015}]{2015AJ....149...77D}
{Drory} N.,  et~al., 2015, \mn@doi [\aj] {10.1088/0004-6256/149/2/77}, \href
  {http://adsabs.harvard.edu/abs/2015AJ....149...77D} {149, 77}

\bibitem[\protect\citeauthoryear{{Dutton} \& {Treu}}{{Dutton} \&
  {Treu}}{2014}]{2014MNRAS.438.3594D}
{Dutton} A.~A.,  {Treu} T.,  2014, \mn@doi [\mnras] {10.1093/mnras/stt2489},
  \href {http://adsabs.harvard.edu/abs/2014MNRAS.438.3594D} {438, 3594}

\bibitem[\protect\citeauthoryear{{Dutton} et~al.,}{{Dutton}
  et~al.}{2011}]{2011MNRAS.417.1621D}
{Dutton} A.~A.,  et~al., 2011, \mn@doi [\mnras]
  {10.1111/j.1365-2966.2011.18706.x}, \href
  {http://adsabs.harvard.edu/abs/2011MNRAS.417.1621D} {417, 1621}

\bibitem[\protect\citeauthoryear{{Eisenstein} et~al.,}{{Eisenstein}
  et~al.}{2001}]{2001AJ....122.2267E}
{Eisenstein} D.~J.,  et~al., 2001, \mn@doi [\aj] {10.1086/323717}, \href
  {http://adsabs.harvard.edu/abs/2001AJ....122.2267E} {122, 2267}

\bibitem[\protect\citeauthoryear{{Eisenstein} et~al.,}{{Eisenstein}
  et~al.}{2011}]{2011AJ....142...72E}
{Eisenstein} D.~J.,  et~al., 2011, \mn@doi [\aj] {10.1088/0004-6256/142/3/72},
  \href {http://adsabs.harvard.edu/abs/2011AJ....142...72E} {142, 72}

\bibitem[\protect\citeauthoryear{{Gunn} et~al.,}{{Gunn}
  et~al.}{2006}]{2006AJ....131.2332G}
{Gunn} J.~E.,  et~al., 2006, \mn@doi [\aj] {10.1086/500975}, \href
  {http://adsabs.harvard.edu/abs/2006AJ....131.2332G} {131, 2332}

\bibitem[\protect\citeauthoryear{{Hewett}, {Warren}, {Willis}, {Bland-Hawthorn}
   \& {Lewis}}{{Hewett} et~al.}{2000}]{2000ASPC..195...94H}
{Hewett} P.~C.,  {Warren} S.~J.,  {Willis} J.~P.,  {Bland-Hawthorn} J.,
  {Lewis} G.~F.,  2000, in {van Breugel} W.,  {Bland-Hawthorn} J.,  eds,
  Astronomical Society of the Pacific Conference Series Vol. 195, Imaging the
  Universe in Three Dimensions. p.~94 (\mn@eprint {} {astro-ph/9905316})

\bibitem[\protect\citeauthoryear{{Hezaveh} et~al.,}{{Hezaveh}
  et~al.}{2016}]{2016ApJ...823...37H}
{Hezaveh} Y.~D.,  et~al., 2016, \mn@doi [\apj] {10.3847/0004-637X/823/1/37},
  \href {http://adsabs.harvard.edu/abs/2016ApJ...823...37H} {823, 37}

\bibitem[\protect\citeauthoryear{{Jim{\'e}nez-Vicente}, {Mediavilla},
  {Kochanek}  \& {Mu{\~n}oz}}{{Jim{\'e}nez-Vicente}
  et~al.}{2015}]{2015ApJ...799..149J}
{Jim{\'e}nez-Vicente} J.,  {Mediavilla} E.,  {Kochanek} C.~S.,   {Mu{\~n}oz}
  J.~A.,  2015, \mn@doi [\apj] {10.1088/0004-637X/799/2/149}, \href
  {http://adsabs.harvard.edu/abs/2015ApJ...799..149J} {799, 149}

\bibitem[\protect\citeauthoryear{{Keeton}, {Kochanek}  \& {Falco}}{{Keeton}
  et~al.}{1998}]{1998ApJ...509..561K}
{Keeton} C.~R.,  {Kochanek} C.~S.,   {Falco} E.~E.,  1998, \mn@doi [\apj]
  {10.1086/306502}, \href {http://adsabs.harvard.edu/abs/1998ApJ...509..561K}
  {509, 561}

\bibitem[\protect\citeauthoryear{{Khostovan}, {Sobral}, {Mobasher}, {Best},
  {Smail}, {Stott}, {Hemmati}  \& {Nayyeri}}{{Khostovan}
  et~al.}{2015}]{2015MNRAS.452.3948K}
{Khostovan} A.~A.,  {Sobral} D.,  {Mobasher} B.,  {Best} P.~N.,  {Smail} I.,
  {Stott} J.~P.,  {Hemmati} S.,   {Nayyeri} H.,  2015, \mn@doi [\mnras]
  {10.1093/mnras/stv1474}, \href
  {http://adsabs.harvard.edu/abs/2015MNRAS.452.3948K} {452, 3948}

\bibitem[\protect\citeauthoryear{{Kochanek}}{{Kochanek}}{1995}]{1995ApJ...445..559K}
{Kochanek} C.~S.,  1995, \mn@doi [\apj] {10.1086/175721}, \href
  {http://adsabs.harvard.edu/abs/1995ApJ...445..559K} {445, 559}

\bibitem[\protect\citeauthoryear{{Koopmans} et~al.,}{{Koopmans}
  et~al.}{2009}]{2009ApJ...703L..51K}
{Koopmans} L.~V.~E.,  et~al., 2009, \mn@doi [\apjl]
  {10.1088/0004-637X/703/1/L51}, \href
  {http://adsabs.harvard.edu/abs/2009ApJ...703L..51K} {703, L51}

\bibitem[\protect\citeauthoryear{{Kroupa}}{{Kroupa}}{2001}]{2001MNRAS.322..231K}
{Kroupa} P.,  2001, \mn@doi [\mnras] {10.1046/j.1365-8711.2001.04022.x}, \href
  {http://adsabs.harvard.edu/abs/2001MNRAS.322..231K} {322, 231}

\bibitem[\protect\citeauthoryear{{Lagattuta}, {Vegetti}, {Fassnacht}, {Auger},
  {Koopmans}  \& {McKean}}{{Lagattuta} et~al.}{2012}]{2012MNRAS.424.2800L}
{Lagattuta} D.~J.,  {Vegetti} S.,  {Fassnacht} C.~D.,  {Auger} M.~W.,
  {Koopmans} L.~V.~E.,   {McKean} J.~P.,  2012, \mn@doi [\mnras]
  {10.1111/j.1365-2966.2012.21406.x}, \href
  {http://adsabs.harvard.edu/abs/2012MNRAS.424.2800L} {424, 2800}

\bibitem[\protect\citeauthoryear{{Law} et~al.,}{{Law}
  et~al.}{2015}]{2015AJ....150...19L}
{Law} D.~R.,  et~al., 2015, \mn@doi [\aj] {10.1088/0004-6256/150/1/19}, \href
  {http://adsabs.harvard.edu/abs/2015AJ....150...19L} {150, 19}

\bibitem[\protect\citeauthoryear{{Law} et~al.,}{{Law}
  et~al.}{2016}]{2016AJ....152...83L}
{Law} D.~R.,  et~al., 2016, \mn@doi [\aj] {10.3847/0004-6256/152/4/83}, \href
  {http://adsabs.harvard.edu/abs/2016AJ....152...83L} {152, 83}

\bibitem[\protect\citeauthoryear{{Linder}}{{Linder}}{2017}]{2017APh....86...41L}
{Linder} E.~V.,  2017, \mn@doi [Astroparticle Physics]
  {10.1016/j.astropartphys.2016.11.002}, \href
  {http://adsabs.harvard.edu/abs/2017APh....86...41L} {86, 41}

\bibitem[\protect\citeauthoryear{{Maraston} \& {Str{\"o}mb{\"a}ck}}{{Maraston}
  \& {Str{\"o}mb{\"a}ck}}{2011}]{2011MNRAS.418.2785M}
{Maraston} C.,  {Str{\"o}mb{\"a}ck} G.,  2011, \mn@doi [\mnras]
  {10.1111/j.1365-2966.2011.19738.x}, \href
  {http://adsabs.harvard.edu/abs/2011MNRAS.418.2785M} {418, 2785}

\bibitem[\protect\citeauthoryear{{Newman}, {Smith}, {Conroy}, {Villaume}  \&
  {van Dokkum}}{{Newman} et~al.}{2017}]{2017ApJ...845..157N}
{Newman} A.~B.,  {Smith} R.~J.,  {Conroy} C.,  {Villaume} A.,   {van Dokkum}
  P.,  2017, \mn@doi [\apj] {10.3847/1538-4357/aa816d}, \href
  {http://adsabs.harvard.edu/abs/2017ApJ...845..157N} {845, 157}

\bibitem[\protect\citeauthoryear{{SDSS Collaboration} et~al.,}{{SDSS
  Collaboration} et~al.}{2017}]{2017arXiv170709322A}
{SDSS Collaboration} et~al., 2017, preprint, \href
  {http://adsabs.harvard.edu/abs/2017arXiv170709322A} {} (\mn@eprint {arXiv}
  {1707.09322})

\bibitem[\protect\citeauthoryear{{Shankar} et~al.,}{{Shankar}
  et~al.}{2017}]{2017ApJ...840...34S}
{Shankar} F.,  et~al., 2017, \mn@doi [\apj] {10.3847/1538-4357/aa66ce}, \href
  {http://adsabs.harvard.edu/abs/2017ApJ...840...34S} {840, 34}

\bibitem[\protect\citeauthoryear{{Shu} et~al.,}{{Shu}
  et~al.}{2015}]{2015ApJ...803...71S}
{Shu} Y.,  et~al., 2015, \mn@doi [\apj] {10.1088/0004-637X/803/2/71}, \href
  {http://adsabs.harvard.edu/abs/2015ApJ...803...71S} {803, 71}

\bibitem[\protect\citeauthoryear{{Shu} et~al.,}{{Shu}
  et~al.}{2016a}]{2016ApJ...824...86S}
{Shu} Y.,  et~al., 2016a, \mn@doi [\apj] {10.3847/0004-637X/824/2/86}, \href
  {http://adsabs.harvard.edu/abs/2016ApJ...824...86S} {824, 86}

\bibitem[\protect\citeauthoryear{{Shu} et~al.,}{{Shu}
  et~al.}{2016b}]{2016ApJ...833..264S}
{Shu} Y.,  et~al., 2016b, \mn@doi [\apj] {10.3847/1538-4357/833/2/264}, \href
  {http://adsabs.harvard.edu/abs/2016ApJ...833..264S} {833, 264}

\bibitem[\protect\citeauthoryear{{Smee} et~al.,}{{Smee}
  et~al.}{2013}]{2013AJ....146...32S}
{Smee} S.~A.,  et~al., 2013, \mn@doi [\aj] {10.1088/0004-6256/146/2/32}, \href
  {http://adsabs.harvard.edu/abs/2013AJ....146...32S} {146, 32}

\bibitem[\protect\citeauthoryear{{Smith}}{{Smith}}{2017}]{2017MNRAS.464L..46S}
{Smith} R.~J.,  2017, \mn@doi [\mnras] {10.1093/mnrasl/slw174}, \href
  {http://adsabs.harvard.edu/abs/2017MNRAS.464L..46S} {464, L46}

\bibitem[\protect\citeauthoryear{{Sonnenfeld}, {Treu}, {Gavazzi}, {Marshall},
  {Auger}, {Suyu}, {Koopmans}  \& {Bolton}}{{Sonnenfeld}
  et~al.}{2012}]{2012ApJ...752..163S}
{Sonnenfeld} A.,  {Treu} T.,  {Gavazzi} R.,  {Marshall} P.~J.,  {Auger} M.~W.,
  {Suyu} S.~H.,  {Koopmans} L.~V.~E.,   {Bolton} A.~S.,  2012, \mn@doi [\apj]
  {10.1088/0004-637X/752/2/163}, \href
  {http://adsabs.harvard.edu/abs/2012ApJ...752..163S} {752, 163}

\bibitem[\protect\citeauthoryear{{Sonnenfeld}, {Gavazzi}, {Suyu}, {Treu}  \&
  {Marshall}}{{Sonnenfeld} et~al.}{2013}]{2013ApJ...777...97S}
{Sonnenfeld} A.,  {Gavazzi} R.,  {Suyu} S.~H.,  {Treu} T.,   {Marshall} P.~J.,
  2013, \mn@doi [\apj] {10.1088/0004-637X/777/2/97}, \href
  {http://adsabs.harvard.edu/abs/2013ApJ...777...97S} {777, 97}

\bibitem[\protect\citeauthoryear{{Spiniello}, {Koopmans}, {Trager}, {Czoske}
  \& {Treu}}{{Spiniello} et~al.}{2011}]{2011MNRAS.417.3000S}
{Spiniello} C.,  {Koopmans} L.~V.~E.,  {Trager} S.~C.,  {Czoske} O.,   {Treu}
  T.,  2011, \mn@doi [\mnras] {10.1111/j.1365-2966.2011.19458.x}, \href
  {http://adsabs.harvard.edu/abs/2011MNRAS.417.3000S} {417, 3000}

\bibitem[\protect\citeauthoryear{{Treu}, {Koopmans}, {Bolton}, {Burles}  \&
  {Moustakas}}{{Treu} et~al.}{2006}]{2006ApJ...640..662T}
{Treu} T.,  {Koopmans} L.~V.,  {Bolton} A.~S.,  {Burles} S.,   {Moustakas}
  L.~A.,  2006, \mn@doi [\apj] {10.1086/500124}, \href
  {http://adsabs.harvard.edu/abs/2006ApJ...640..662T} {640, 662}

\bibitem[\protect\citeauthoryear{{Treu}, {Dutton}, {Auger}, {Marshall},
  {Bolton}, {Brewer}, {Koo}  \& {Koopmans}}{{Treu}
  et~al.}{2011}]{2011MNRAS.417.1601T}
{Treu} T.,  {Dutton} A.~A.,  {Auger} M.~W.,  {Marshall} P.~J.,  {Bolton} A.~S.,
   {Brewer} B.~J.,  {Koo} D.~C.,   {Koopmans} L.~V.~E.,  2011, \mn@doi [\mnras]
  {10.1111/j.1365-2966.2011.19378.x}, \href
  {http://adsabs.harvard.edu/abs/2011MNRAS.417.1601T} {417, 1601}

\bibitem[\protect\citeauthoryear{{Vegetti} \& {Koopmans}}{{Vegetti} \&
  {Koopmans}}{2009}]{Vegetti09a}
{Vegetti} S.,  {Koopmans} L.~V.~E.,  2009, \mn@doi [\mnras]
  {10.1111/j.1365-2966.2008.14005.x}, \href
  {http://adsabs.harvard.edu/abs/2009MNRAS.392..945V} {392, 945}

\bibitem[\protect\citeauthoryear{{Vegetti}, {Koopmans}, {Bolton}, {Treu}  \&
  {Gavazzi}}{{Vegetti} et~al.}{2010}]{2010MNRAS.408.1969V}
{Vegetti} S.,  {Koopmans} L.~V.~E.,  {Bolton} A.,  {Treu} T.,   {Gavazzi} R.,
  2010, \mn@doi [\mnras] {10.1111/j.1365-2966.2010.16865.x}, \href
  {http://adsabs.harvard.edu/abs/2010MNRAS.408.1969V} {408, 1969}

\bibitem[\protect\citeauthoryear{{Vegetti}, {Lagattuta}, {McKean}, {Auger},
  {Fassnacht}  \& {Koopmans}}{{Vegetti} et~al.}{2012}]{2012Natur.481..341V}
{Vegetti} S.,  {Lagattuta} D.~J.,  {McKean} J.~P.,  {Auger} M.~W.,  {Fassnacht}
  C.~D.,   {Koopmans} L.~V.~E.,  2012, \mn@doi [\nat] {10.1038/nature10669},
  \href {http://adsabs.harvard.edu/abs/2012Natur.481..341V} {481, 341}

\bibitem[\protect\citeauthoryear{{Vegetti}, {Koopmans}, {Auger}, {Treu}  \&
  {Bolton}}{{Vegetti} et~al.}{2014}]{2014MNRAS.442.2017V}
{Vegetti} S.,  {Koopmans} L.~V.~E.,  {Auger} M.~W.,  {Treu} T.,   {Bolton}
  A.~S.,  2014, \mn@doi [\mnras] {10.1093/mnras/stu943}, \href
  {http://adsabs.harvard.edu/abs/2014MNRAS.442.2017V} {442, 2017}

\bibitem[\protect\citeauthoryear{{Wake} et~al.,}{{Wake}
  et~al.}{2017}]{2017AJ....154...86W}
{Wake} D.~A.,  et~al., 2017, \mn@doi [\aj] {10.3847/1538-3881/aa7ecc}, \href
  {http://adsabs.harvard.edu/abs/2017AJ....154...86W} {154, 86}

\bibitem[\protect\citeauthoryear{{Warren}, {Hewett}, {Lewis}, {M{\o}ller},
  {Iovino}  \& {Shaver}}{{Warren} et~al.}{1996}]{1996MNRAS.278..139W}
{Warren} S.~J.,  {Hewett} P.~C.,  {Lewis} G.~F.,  {M{\o}ller} P.,  {Iovino} A.,
    {Shaver} P.~A.,  1996, \mn@doi [\mnras] {10.1093/mnras/278.1.139}, \href
  {http://adsabs.harvard.edu/abs/1996MNRAS.278..139W} {278, 139}

\bibitem[\protect\citeauthoryear{{Wilkinson} et~al.,}{{Wilkinson}
  et~al.}{2015}]{2015MNRAS.449..328W}
{Wilkinson} D.~M.,  et~al., 2015, \mn@doi [\mnras] {10.1093/mnras/stv301},
  \href {http://adsabs.harvard.edu/abs/2015MNRAS.449..328W} {449, 328}

\bibitem[\protect\citeauthoryear{{Willis}, {Hewett}  \& {Warren}}{{Willis}
  et~al.}{2005}]{2005MNRAS.363.1369W}
{Willis} J.~P.,  {Hewett} P.~C.,   {Warren} S.~J.,  2005, \mn@doi [\mnras]
  {10.1111/j.1365-2966.2005.09533.x}, \href
  {http://adsabs.harvard.edu/abs/2005MNRAS.363.1369W} {363, 1369}

\bibitem[\protect\citeauthoryear{{Willis}, {Hewett}, {Warren}, {Dye}  \&
  {Maddox}}{{Willis} et~al.}{2006}]{2006MNRAS.369.1521W}
{Willis} J.~P.,  {Hewett} P.~C.,  {Warren} S.~J.,  {Dye} S.,   {Maddox} N.,
  2006, \mn@doi [\mnras] {10.1111/j.1365-2966.2006.10399.x}, \href
  {http://adsabs.harvard.edu/abs/2006MNRAS.369.1521W} {369, 1521}

\bibitem[\protect\citeauthoryear{{Yan} et~al.,}{{Yan}
  et~al.}{2016a}]{2016AJ....151....8Y}
{Yan} R.,  et~al., 2016a, \mn@doi [\aj] {10.3847/0004-6256/151/1/8}, \href
  {http://adsabs.harvard.edu/abs/2016AJ....151....8Y} {151, 8}

\bibitem[\protect\citeauthoryear{{Yan} et~al.,}{{Yan}
  et~al.}{2016b}]{2016AJ....152..197Y}
{Yan} R.,  et~al., 2016b, \mn@doi [\aj] {10.3847/0004-6256/152/6/197}, \href
  {http://adsabs.harvard.edu/abs/2016AJ....152..197Y} {152, 197}

\bibitem[\protect\citeauthoryear{{York} et~al.,}{{York}
  et~al.}{2000}]{2000AJ....120.1579Y}
{York} D.~G.,  et~al., 2000, \mn@doi [\aj] {10.1086/301513}, \href
  {http://adsabs.harvard.edu/abs/2000AJ....120.1579Y} {120, 1579}

\bibitem[\protect\citeauthoryear{{van Dokkum} et~al.,}{{van Dokkum}
  et~al.}{2017}]{2017ApJ...844L..11V}
{van Dokkum} P.,  et~al., 2017, \mn@doi [\apjl] {10.3847/2041-8213/aa7ca2},
  \href {http://adsabs.harvard.edu/abs/2017ApJ...844L..11V} {844, L11}

\makeatother
\end{thebibliography}
\end{document}